\documentclass[11pt]{article}
\usepackage[final]{acl}

\usepackage{times}
\usepackage{latexsym}

\usepackage[T1]{fontenc}

\usepackage[utf8]{inputenc}

\usepackage{microtype}

\usepackage{inconsolata}

\usepackage{graphicx}

\usepackage{amsmath}
\usepackage{amsthm}

\usepackage{array}
\usepackage{amsfonts} 
\usepackage{multirow}
\usepackage{adjustbox}
\usepackage{booktabs}
\usepackage{paralist}
\usepackage[ruled,vlined]{algorithm2e}
\usepackage{titletoc}

\usepackage[many]{tcolorbox}
\tcbuselibrary{listings, breakable, skins}

\newtcblisting{promptbox}{
    listing only,
    listing options={
        basicstyle=\ttfamily\footnotesize,
        breaklines=true,
        columns=fullflexible
    },
    colback=gray!5,
    colframe=gray!70,
    boxrule=0.5pt,
    arc=2pt,
    left=4pt,
    right=4pt,
    top=2pt,
    bottom=2pt,
    grow to left by=10pt,
    breakable
}

\title{DarwinTOD: LLM-Driven Lifelong Self-evolution for Task-oriented Dialog Systems}

\author{%
    Shuyu Zhang\textsuperscript{\rm 1},\;
    Yujie Liu\textsuperscript{\rm 2},\;
    Xinru Wang\textsuperscript{\rm 3},\;\\
    \textbf{Cheng Zhang}\textsuperscript{\rm 4},\; 
    \textbf{Yanmin Zhu}\textsuperscript{\rm 1}\thanks{Corresponding Author.}, \; 
    \textbf{Bin Li}\textsuperscript{\rm 5 *}\; 
    \\
    \textsuperscript{\rm 1}Shanghai Jiao Tong University, 
    \textsuperscript{\rm 2}Beijing Institute of Graphic Communication, \\
    \textsuperscript{\rm 3}University of Sydney, 
    \textsuperscript{\rm 4}Tianjin University of Finance and Economics, 
    \textsuperscript{\rm 5}SIAT, CAS \\ 
    \small{
   \href{mailto:carsonz@sjtu.edu.cn}{carsonz@sjtu.edu.cn},
   \href{mailto:yzhu@cs.sjtu.edu.cn}{yzhu@cs.sjtu.edu.cn},
   \href{mailto:b.li2@siat.ac.cn}{b.li2@siat.ac.cn},
 }
}

\begin{document}
\maketitle
\begin{abstract}
Traditional task-oriented dialog systems are unable to evolve from ongoing interactions or adapt to new domains after deployment, that is a critical limitation in real-world dynamic environments. Continual learning approaches depend on episodic retraining with human-curated data, failing to achieve autonomy lifelong improvement. While evolutionary computation and LLM driven self-improvement offer promising mechanisms for dialog optimization, they lack a unified framework for holistic, iterative strategy refinement. To bridge this gap, we propose DarwinTOD\footnote{https://github.com/carsonz/DarwinTOD}, a lifelong self-evolving dialog framework that systematically integrates these two paradigms, enabling continuous strategy optimization from a zero-shot base without task-specific fine-tuning. DarwinTOD maintains an Evolvable Strategy Bank and operates through a dual-loop process: online multi-agent dialog execution with peer critique, and offline structured evolutionary operations that refine the strategy bank using accumulated feedback. This closed-loop design enables autonomous continuous improvement without human intervention. Extensive experiments show that DarwinTOD surpasses previous state-of-the-art methods and exhibits continuous performance gains throughout evolution. Our work provides a novel framework for building dialog systems with lifelong self-evolution capabilities.
\end{abstract}

\section{Introduction}

\begin{figure*}[t!]
    \centering
    \includegraphics[width=1\linewidth]{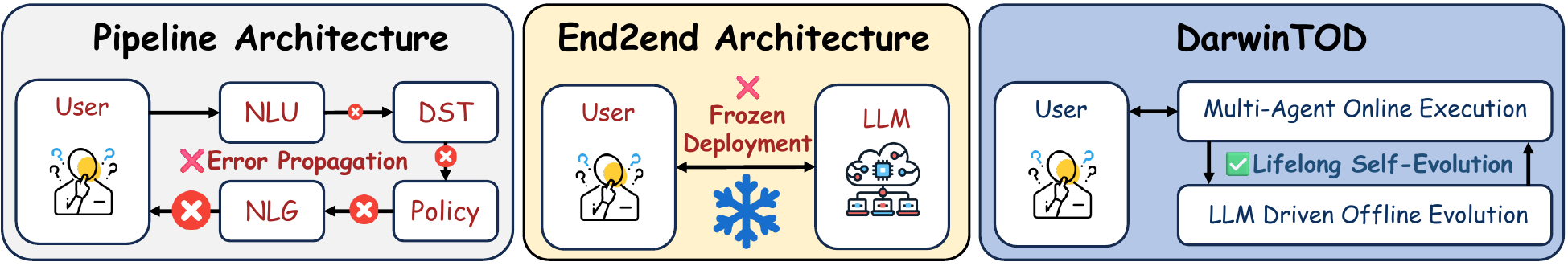}
    \caption{Motivation comparison of TOD architectures. Both pipeline and end-to-end TOD systems suffer from cascaded errors or lack experience driven improvement, while DarwinTOD enables lifelong self-evolution via a dual-loop architecture to achieve autonomous improvement.}
    \label{fig:motivation}
\end{figure*}

Task-oriented dialog (TOD) systems aim to assist users in achieving specific goals through natural language conversations, with applications like customer service and personal assistants. Despite advances, deployed systems remain static, unable to learn from ongoing interactions~\citep{madotto-etal-2021-continual}. This conflicts with real-world needs where user preferences and domains evolve continuously. Truly intelligent dialog agents require lifelong self-evolution capabilities to refine strategies throughout their operational lifespan. This creates a critical gap between research prototypes and deployable systems: the former are often evaluated on static benchmarks, while the latter must operate in a dynamic, open-ended world. Bridging this gap necessitates a paradigm shift towards systems endowed with lifelong self-evolution to autonomously refine their conversational strategies through continuous interaction.

Current TOD paradigms fail to meet this requirement. Pipeline modular systems decompose dialog into components like natural language understanding (NLU), dialog state tracking (DST), dialog policy (DP) and natural language generation (NLG)~\citep{DING2024111662, GONG2025112754}, but suffer from cascaded error propagation and become brittle to new domains~\citep{10.1007/s00521-023-09328-9}. End-to-end LLM-based approaches show strong generalization via instruction following~\citep{king-flanigan-2024-unsupervised, xu-etal-2024-rethinking}, but remain static after initial training, lacking continuous improvement mechanisms~\citep{LI2025130991}. Even continual learning methods rely on episodic retraining with curated data~\citep{zeng-etal-2025-task, Xu_Yang_Xu_2023}, not achieving autonomous evolution. The emergence of modern large language models (LLMs) provides a new foundation with their advanced instruction following, reasoning, and text generation capabilities. As depicted in Figure~\ref{fig:motivation}, these limitations motivate our core \textbf{Research Question}: How can we achieve lifelong self-evolution for TOD system, enabling it to continuously and autonomously improve from its own interactions?

Evolutionary computation and LLM driven self-improvement offer promising directions. Evolutionary algorithms enable population-based optimization for prompts~\citep{10.5555/3692070.3692611, ICLR2024_9156b0f6} and policies~\citep{zhao-etal-2025-efficient}. LLM-based multi-agent systems solve complex problems through collaboration~\citep{su-etal-2025-many, Cheng_Liu_Wang_Leong_Ouyang_Li_Wu_Zheng_2024}, while self-evolving frameworks refine agent behavior through experience~\citep{zhang2025memgenweavinggenerativelatent, fang2025comprehensivesurveyselfevolvingai}. However, these approaches remain fragmented: evolutionary methods optimize isolated prompts without addressing holistic dialog strategy lifecycles; multi-agent systems often focus on single-turn dialog; and self-evolving frameworks lack structured dialog management.

We introduce DarwinTOD, a lifelong self-evolving dialog framework that integrates evolutionary computation with modern LLM driven strategy optimization. Unlike conventional prompt engineering, DarwinTOD operates as a population-based evolutionary system that undergoes parallel competition, fitness based selection, and elimination through a closed-loop, lifelong evolutionary cycle. Its core is an Evolvable Strategy Bank (ESB) and a dual-loop process: online multi-agent execution with peer critique, followed by offline structured evolution using accumulated feedback. This closed-loop design enables fully autonomous self-improvement from a minimal starting point, without task-specific fine-tuning or human curation. Our contributions are as follows. 

\noindent \textbf{(1)} We introduce a novel lifelong self-evolution framework for TOD that systematically integrates LLM-driven evolutionary optimization.

\noindent \textbf{(2)} We propose a structured mechanism centered on dynamic ESB and dual-loop cycle that enables self-improvement without human intervention.

\noindent \textbf{(3)} Extensive empirical validation showing state-of-the-art (SOTA) performance through sustained autonomous evolution.

\section{Related Work}

\textbf{TOD Systems.}
TOD research has evolved through distinct paradigms, each addressing aspects of generalization and adaptability, but none achieving true lifelong autonomy. Pipeline systems decompose dialog into specialized components, enabling interpretability but suffering from error propagation and costly domain re-engineering~\citep{wu-etal-2019-transferable, Zhang_Ou_Yu_2020}. End-to-end approaches leverage LLMs to generate responses directly, improving generalization via instruction following~\citep{10.5555/3495724.3497418, Yang_Li_Quan_2021} but remaining fixed after deployment. Continual learning methods introduce incremental updates to handle new domains or tasks~\citep{Liu_Mazumder_2021, madotto-etal-2021-continual, zhao-etal-2022-prompt, kimMeDiTODERMedicalDomainincremental2025}, yet operate with explicit task boundaries and require episodic retraining with curated data or generative replay. Recent works have highlighted the need for continuous learning during conversation~\citep{10.1145/3477495.3532677} and self-evolution capabilities~\citep{tao2024surveyselfevolutionlargelanguage}. A parallel work explores LLM-based agents that can call external tools or APIs dynamically~\citep{xu-etal-2024-rethinking, 10.1145/3745021}, and self explanation prompting can improve dialog understanding~\citep{gao-etal-2024-self}, but these methods remain static after deployment. These limitations collectively highlight the need for a paradigm shift toward systems capable of endogenous self-evolution without human intervention.

\textbf{Evolutionary Computation.}
Evolutionary algorithms provide a population-based optimization paradigm, while LLMs with remarkable prowess as general-purpose controllers and optimizers serve as intelligent evolution operators. Classical EAs have been applied to TOD \citep{zhao-etal-2025-efficient} but lack semantic awareness. LLM driven evolution shows promise in prompt engineering \citep{10.5555/3692070.3692611, agarwal-etal-2025-promptwizard} and game generation \citep{10.5555/3737916.3741431}, with surveys outlining this synergy \citep{10767756}. However, existing methods target single-turn or static-goal tasks, lacking systematic support for multi-turn dialog challenges within a lifelong learning framework.

\textbf{Self Evolving Agents.}
Research on self-improving agents spans self-reflection \citep{ICLR2025_afb27164}, multi-agent collaboration \citep{su-etal-2025-many, chen-etal-2025-optima}, and memory-augmented learning \citep{tan-etal-2025-prospect}. Recent frameworks enable iterative self-improvement via LLMs \citep{zhang2025marsmultiagentadaptivereasoning, zhai2025agentevolverefficientselfevolvingagent, gao-etal-2024-self-evolving}. However, these approaches typically focus on updating parameters of a single agent or optimizing for single-turn tasks, and do not address the unique challenges of multi-domain and multi-turn conversational strategy evolution in TOD. DarwinTOD addresses these gaps by introducing a dedicated ESB as an evolving population, dialog specific evolutionary operators, and a dual-loop architecture for continuous autonomous refinement.

\begin{figure*}[t]
    \centering
    \includegraphics[width=1\linewidth]{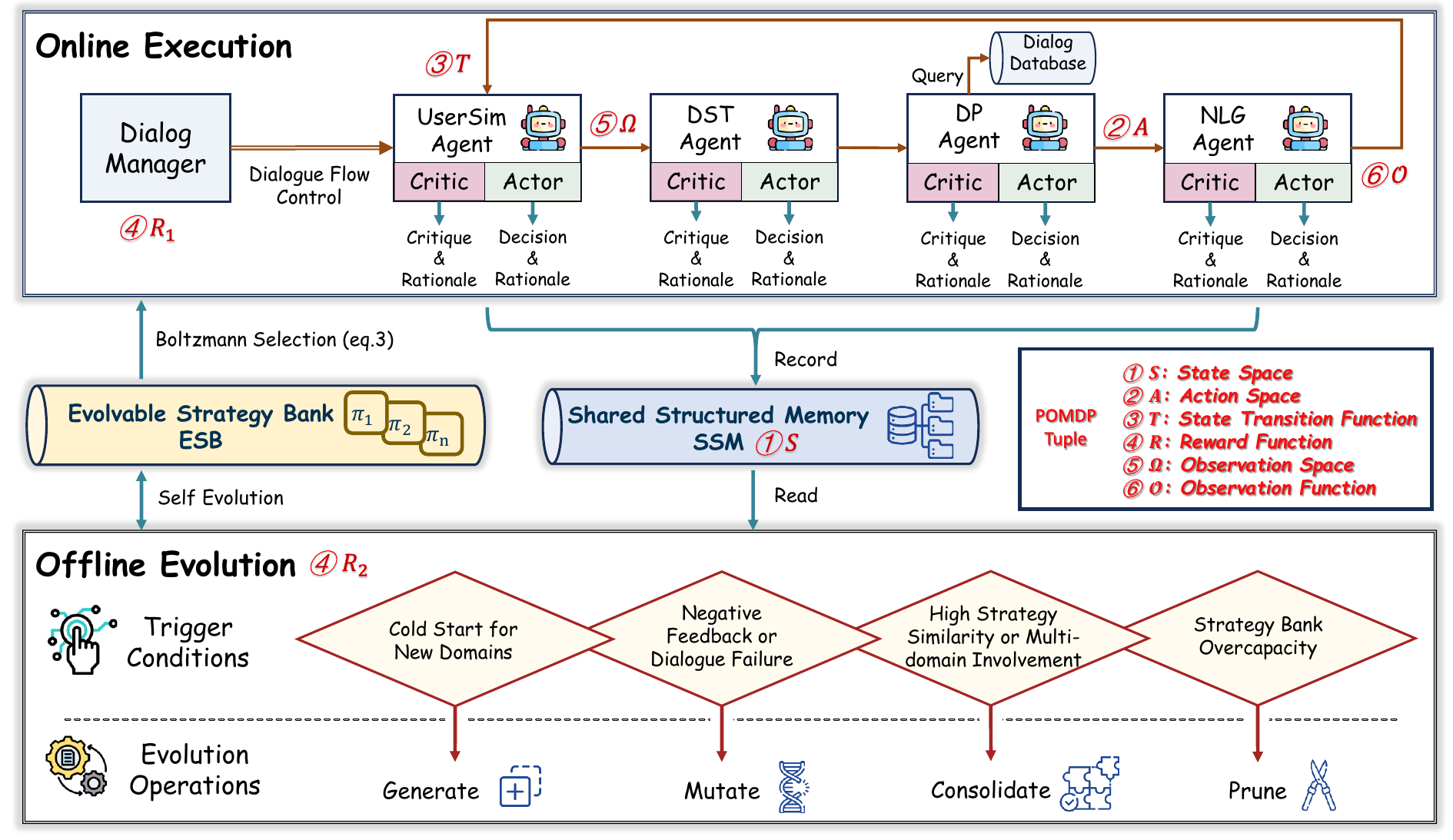}
    \caption{DarwinTOD's dual-loop algorithm framework. The online phase executes dialogs via multi-agent collaboration (DST/DP/NLG/UserSim) with peer critique, retrieving strategies from ESB through Boltzmann selection and logging interactions to SSM. The offline phase triggers evolutionary operations (Generate/Mutate/Consolidate/Prune) based on SSM feedback to update ESB, forming a closed-loop for autonomous strategy refinement.}
    \label{fig:flowchart}
\end{figure*}

\section{Methodology}
\label{sec:methodology}

DarwinTOD unifies evolutionary computation with LLM driven strategy optimization, moving beyond single prompt tuning to a population-based evolutionary paradigm where strategies compete, mutate, and are selected through a structured dual-loop process. We establish its theoretical foundation by formalizing dialog as a Partially Observable Markov Decision Process (POMDP) and strategy evolution as a Markov chain, instantiated in a dual-loop algorithmic framework that enables autonomous lifelong adaptation without human intervention.

\subsection{Theoretical Foundation}
\label{subsec:theoretical_foundation}

We formalize TOD as a POMDP to capture its sequential and partially observable nature, defined by the tuple $\{\mathcal{S},\mathcal{A},\mathcal{T},\mathcal{R},\Omega,\mathcal{O}\}$. Here, $\mathcal{S}$ is the state space, typically comprising the user goal $g$, database results $db$, and dialog history $h$; $\mathcal{A}$ is the action space of system acts; $\mathcal{T}$ is the state transition function, implemented by the user simulator; $\mathcal{R}$ is the reward function, provided by the dialog manager and offline evolution; $\Omega$ is the observation space (user utterance $u$); $\mathcal{O}$ is the observation function, corresponding to user utterance generation. The system maintains a belief state $b_t(s)$. Our objective is to optimize a dialog policy $\pi: b \mapsto a$ that maps belief states to actions, maximizing the expected cumulative reward:
\begin{equation}
\label{eq:reward}
J(\pi) = \mathbb{E}_{\zeta \sim p(\zeta|\pi)} \left[ \sum_{t=0}^{T} \mathcal{R}(s_{t},a_{t}) \right].
\end{equation}
where $\zeta$ denotes a dialog trajectory.

We model the lifelong evolution of dialog strategies by treating the Evolvable Strategy Bank (ESB) as a population in a Markov chain over generations. The ESB in generation $t$, denoted $\Pi_t$, transitions to $\Pi_{t+1}$ through selection, feedback evaluation, and evolutionary operators. Each strategy $\pi \in \Pi_t$ is assigned a fitness score $\phi(\pi)$ that balances its historical performance against its age to prevent stagnation:
\begin{equation}
\label{eq:fitness}
\phi(\pi) = \frac{H_\pi^{+} - H_\pi^{-}}{N_\pi + \epsilon} + \alpha \cdot \text{norm}(\pi_\text{gen}).
\end{equation}
Here, $H_\pi^{+}$/$H_\pi^{-}$ are positive/negative feedback counts (updated after each dialog based on task success and peer critiques), $N_\pi$ is the total usage count, $\pi_\text{gen}$ is the generation index, and $\text{norm}(\cdot)$ denotes global min-max normalization to scale $\pi_\text{gen}$ to the range $[0, 1]$. The term $\epsilon$ is a smoothing constant and $\alpha$ controls the age penalty, which discourages the use of older strategies and prevents premature convergence. This fitness function explicitly balances exploitation of high-performing strategies with penalization of older strategies, similar to evolutionary algorithms that maintain diversity and avoid premature convergence.

During online execution, a strategy $\pi_i$ applicable to domain $d$ is selected stochastically according to a Boltzmann distribution based on its fitness:
\begin{equation}
\label{eq:Boltzmann_selection}
P(\pi_i | d) = \frac{\mathbb{I}(d \in d_i) \cdot \exp\big(\phi(\pi_i)/\tau\big)}
{\sum_j \mathbb{I}(d \in d_j) \cdot \exp\big(\phi(\pi_j)/\tau\big)}.
\end{equation}
where $\mathbb{I}(\cdot)$ is the indicator function, $d_i$ is the set of domains, this function restricts selection to strategies applicable to the current domains $d$. And $\tau > 0$ is a temperature parameter controlling the exploration exploitation trade-off, a higher $\tau$ encourages exploration of lower fitness strategies, while a lower $\tau$ favors exploitation of high-fitness ones, analogous to simulated annealing in evolutionary computation.

As detailed analysis in Appendix~\ref{app:complete_theorem}, the evolutionary process is designed to be robust to noisy mutations and potential biases in LLM generated critiques. The fitness function (Eq.~\ref{eq:fitness}) smooths and normalizes feedback over multi-turn interactions, guiding selection pressure toward strategies with consistently high performance. Boltzmann selection (Eq.~\ref{eq:Boltzmann_selection}) balances exploration and exploitation based on long term fitness, preventing premature convergence to local optima caused by single noisy critiques. This structured evolution constitutes a directed search in semantic strategy space, where periodic pruning eliminates persistently low-fitness candidates. Consequently, the closed-loop design enables the ESB to progressively concentrate on higher performing strategies over time, as validated by experiments in Sec.~\ref{sec:main_results}.

\subsection{Algorithmic Framework}
\label{sec:algorithmic_framework}

Figure~\ref{fig:flowchart} depicts the dual-loop architecture of DarwinTOD, which couples online POMDP evaluation with offline Markov chain based evolution. This materializes the theory analysis into an executable algorithmic process. The complete pseudocode and prompts are provided in Sec.~\ref{app:algorithm_pseudocode} and~\ref{app:all_prompts} respectively.

\subsubsection{Core Data Structures}

The ESB maintains a population of dialog strategies, each represented as a tuple $\pi_i = \{id_i, d_i, c_i, m_i\}$. Here, $id_i$ is a unique identifier, $d_i$ denotes applicable domains, $c_i$ is the natural language description of the strategy, and $m_i = \{h_i^+, h_i^-, n_i, \text{gen}_i\}$ records metadata: positive/negative feedback counts, usage count, and generation index. Each strategy also includes a rationale field and a lifecycle flag indicating whether it is active.

The SSM stores complete dialog trajectories for evolutionary learning. Each trajectory $\mathcal{H}$ is structured as a tuple $\{d, g, \Pi^{\text{used}}, { (u_t, r_t, b_t, a_t, c_t)}_{t=1}^T \}$, where $d$ is domain/domains, $g$ the user goal, $\Pi^{\text{used}}$ is the set of strategies employed, and each turn contains the user utterance $u_t$, system response $r_t$, the current belief state $b_t$, the system action $a_t$, and a structured critique log $c_t$, which includes rationales for each agent and self evaluations, which are used during offline evolution.

ESB and SSM instantiate the theoretical strategy population and experience buffer, respectively.

\subsubsection{Online Execution}

The online phase implements the POMDP policy evaluation through four specialized LLM agents: DST (Dialog State Tracker), DP (Dialog Policy), NLG (Natural Language Generator), and UserSim (User Simulator). For each dialog turn $t$:

\noindent \textbf{Strategy Retrieval}: Each agent retrieves a strategy from ESB using the domain-aware Boltzmann selection defined in Eq.~\ref{eq:Boltzmann_selection}.

\noindent \textbf{Multi-Agent Execution}: Agents operate sequentially with built-in critique:
    \[
    u_t \xrightarrow[\text{critique}]{\text{DST}} b_t 
    \xrightarrow[\text{critique}]{\text{DP}} a_t 
    \xrightarrow[\text{critique}]{\text{NLG}} r_t
    \xrightarrow[\text{critique}]{\text{UserSim}}.
    \]
Each agent first critiques the previous agent's output, and then produces its own output with justification. This pipeline realizes POMDP evaluation (Eq.~\ref{eq:reward}): DST updates the belief state $b_t$, DP selects the system action $a_t$ and queries database if needed, NLG generates the response $r_t$, and UserSim provides critique based on $r_t$. These critiques serve as immediate reward signals for offline evolution.

\noindent \textit{Note}: In the main experiments, UserSim does not generate the next user utterance $u_{t+1}$; instead, it leverages the next user utterance from the dataset.

\noindent \textbf{Feedback Collection}: Dialog Manager provides the task success assessment; the usage count $n_i$ of each involved strategy is updated; and the entire interaction record and agent critiques are stored in SSM. This collected experience serves as the training corpus for offline strategy evolution.

\subsubsection{Offline Evolution}
After each dialog episode, the offline phase refines ESB through four evolutionary operators, triggered by interaction feedback and population metrics. The system sustains autonomous improvement amid noisy variation, and not depend on every critique or every mutation. Instead, by selecting and pruning, it filters out detrimental changes while preserving productive ones.

\noindent\textbf{(1) Genesis} is triggered when a domain or a combination of domains is encountered and no strategies exist for it. For each agent type, it synthesizes $K$ strategies based solely on the domain name and the agent's role description, without leveraging any prior dialog history. Formally, for a given agent type $ag$ and domain $d$, the operator generates a set of new strategies $G(\Pi_{new}, d) \sim \mathcal{M}_{\text{LLM}}(\text{GEN} \mid d, ag)_{i=1}^{K}$. The resulting strategies are added to ESB, providing an initial policy repertoire for the new domain without any training data in the domain.

\noindent\textbf{(2) Mutation} is applied to strategy $\pi$ that was involved in a failed dialog or received negative critiques. The operator first uses an LLM to assess whether $\pi$ was helpful, neutral, or harmful in the given context, and updates its corresponding feedback counts in the ESB. Then it prompts one to generate a revised strategy $\pi'$ that addresses the identified shortcomings, with the failed trajectory and critiques: $\pi' \sim \mathcal{M}_{\text{LLM}}(\text{MUT} \mid \pi, \mathcal{H})$. The newly created strategy $\pi'$ inherits the metadata of $\pi$ but with its generation index incremented.

\noindent\textbf{(3) Consolidation} merges a set of $n$ highly similar strategies, if the cosine distances of their SBERT encoded strategy texts exceed threshold $\delta$. The new strategy $\pi_c$ is synthesized by prompting an LLM with the combined content of $n$ source strategies, its metadata is set to the average of the original, and the maximum generation index plus one: $\pi_c \sim \mathcal{M}_{\text{LLM}}(\text{CON} \mid \pi_1, \pi_2, \dots, \pi_n)$. The original $n$ strategies are then removed from the ESB to maintain a compact and diverse population.

\noindent\textbf{(4) Pruning} maintains a bounded population size $M$ by discarding the lowest fitness strategies after each episode. Strategies are ranked by fitness $\phi(\pi)$, and only the top-$M$ are retained: $P(\Pi_t) = \Pi_t \setminus \{\pi \mid \text{rank}(\phi(\pi)) > M\}$. This ensures computational efficiency while preserving high-performing and diverse strategies.

\section{Experiments}

\begin{table*}[t]
\caption{Performance comparison between DarwinTOD and baseline models on MultiWOZ 2.0, 2.1, and 2.2. Metrics include: Inform, Success (Succ.), BLEU, and Combine (Comb.). Bold indicates the best score for metrics. All results of baselines were reported from original papers.}
\label{tab:main_result}
\centering
\begin{adjustbox}{max width=\textwidth}
\begin{tabular}{l|c|cccc|cccc|cccc}
\hline
\multirow{2}{*}{Model} & \multirow{2}{*}{Year} & \multicolumn{4}{c|}{MultiWOZ 2.0} & \multicolumn{4}{c|}{MultiWOZ 2.1} & \multicolumn{4}{c}{MultiWOZ 2.2} \\
\cline{3-14}
 & & Inform & Succ. & BLEU & Comb. & Inform & Succ. & BLEU & Comb. & Inform & Succ. & BLEU & Comb. \\
\hline
SimpleTOD & 2020 & 84.40 & 70.10 & 15.01 & 92.26 & 85.00 & 70.50 & 15.23 & 92.98 & - & - & - & - \\
MinTL & 2020 & 84.88 & 74.91 & 17.89 & 97.79 & - & - & - & - & 73.7 & 65.4 & 19.4 & 89.0 \\
SOLOIST & 2021 & 85.50 & 72.90 & 16.54 & 95.74 & - & - & - & - & 82.3 & 72.4 & 13.6 & 90.9 \\
UBAR & 2021 & 95.40 & 80.70 & 17.00 & 105.05 & 95.70 & 81.80 & 16.50 & 105.25 & 83.4 & 70.3 & 17.6 & 94.4 \\
PPTOD & 2022 & 89.20 & 79.40 & 18.62 & 102.92 & 87.09 & 79.08 & 19.17 & 102.26 & - & - & - & - \\
BORT & 2022 & 93.80 & 85.80 & 18.50 & 108.30 & - & - & - & - & 85.5 & 77.4 & 17.9 & 99.4 \\
GALAXY & 2022 & 94.40 & 85.30 & 20.50 & 110.35 & 95.30 & 86.20 & 20.01 & 110.76 & 85.4 & 75.7 & 19.6 & 100.2 \\
TOATOD & 2023 & - & - & - & - & 97.00 & 87.40 & 17.12 & 109.32 & 90.0 & 79.8 & 17.0 & 101.9 \\
SGP-TOD & 2023 & 83.88 & 69.87 & 9.09 & 85.97 & - & - & - & - & 82.00 & 72.50 & 9.22 & 86.47 \\
AutoTOD & 2024 & 91.70 & 84.40 & 10.40 & 98.50 & - & - & - & - & - & - & - & -\\
KMc-ToD & 2024 & 93.90 & 84.70 & 20.20 & 109.30 & 93.20 & 84.90 & 19.60 & 108.70 & - & - & - & -\\
ProTOD & 2025 & 91.70 & 83.30 & 8.90 & 96.40 & - & - & - & - & - & - & - & - \\
OmniDialog & 2025 & 92.20 & 79.30 & 18.57 & 104.32 & - & - & - & - & - & - & - & - \\
AgentTOD (Prev. SOTA) & 2025 & 95.27 & 88.95 & 21.77 & 113.88 & 96.73 & 89.92 & 20.54 & 113.87 & 90.90 & 81.25 & 18.49 & 104.56 \\
\hline
DarwinTOD (Llama3-8B) & 2026& 96.92 & 89.14 & 21.83 & 114.86 & 98.73 & 91.42 & 19.96 & 115.04 & 92.58 & 83.97 & 17.98 & 106.26 \\
DarwinTOD (Qwen2.5-7B) & 2026& 97.63 & 90.28 & 21.55 & 115.51 & 98.92 & 91.85 & 20.18 & 115.57 & 92.14 & 84.33 & 18.34 & 106.58 \\
DarwinTOD (Qwen3-8B) & 2026& 98.34 & 92.86 & 21.74 & 117.34 & 99.62 & 94.18 & 20.33 & 117.23 & 94.73 & 87.25 & 18.41 & 109.40 \\
DarwinTOD (GPT-5.1) & 2026 & \textbf{99.10} & \textbf{96.20} & \textbf{22.94} & \textbf{120.59} & \textbf{99.40} & \textbf{96.50} & \textbf{22.19} & \textbf{120.14} & \textbf{96.48} & \textbf{90.12} & \textbf{21.98} & \textbf{115.28} \\
\hline
\end{tabular}
\end{adjustbox}
\end{table*}

\subsection{Experimental Setup}

\noindent \textbf{Datasets.}
We evaluate DarwinTOD on two established benchmarks: three versions of MultiWOZ~\citep{budzianowski-etal-2018-multiwoz, eric-etal-2020-multiwoz, zang-etal-2020-multiwoz} and SGD~\citep{Rastogi_Zang_Sunkara_Gupta_Khaitan_2020}. These datasets provide multi-domain conversational data and are standard for evaluating TOD system capabilities, details in Appendix~\ref{app:datasets_details}.

\noindent\textbf{Baselines.}
We compare against several strong and recent baselines, including representative pipeline and end-to-end approaches. All baseline results were reported from the original papers. See Appendix~\ref{app:baselines_details} for details.

\noindent \textbf{Metrics.}
Following standard TOD evaluation protocols, we employ four automatic metrics, the same as in the previous SOTA AgentTOD~\citep{xuAgentTODTaskorientedDialog2025}.
\textit{Inform}: Measures whether the system provides the correct entity requested by the user.
\textit{Success}: Measures whether the system successfully answers all user constraints.
\textit{BLEU}~\citep{papineni-etal-2002-bleu}: Evaluates the quality of the generated response comparing the ground truth, assessing fluency.
\textit{Combine}: A composite score defined as $\text{Combine} = (\text{Inform} + \text{Success}) \times 0.5 + \text{BLEU}$, providing an overall performance measure.
For a detailed discussion on the suitability and limitations of BLEU, refer to Appendix~\ref{app:metrics_discussion}.

\noindent \textbf{Implementation Details.}
DarwinTOD reads the user's utterance from the training dataset for online execution, records the process data in SSM, and performs offline evolution after each dialog episode. DarwinTOD starts with initial ESB generated by genesis operation solely on domain names and agent role descriptions. We use Llama, Qwen, and GPT as backbone LLMs. More details, hyperparameters, and prompts are provided in Appendix~\ref{app:detailed_setup} and~\ref{app:all_prompts}.

\subsection{Main Results}
\label{sec:main_results}

\begin{figure}[t]
    \centering
    \includegraphics[width=1\linewidth]{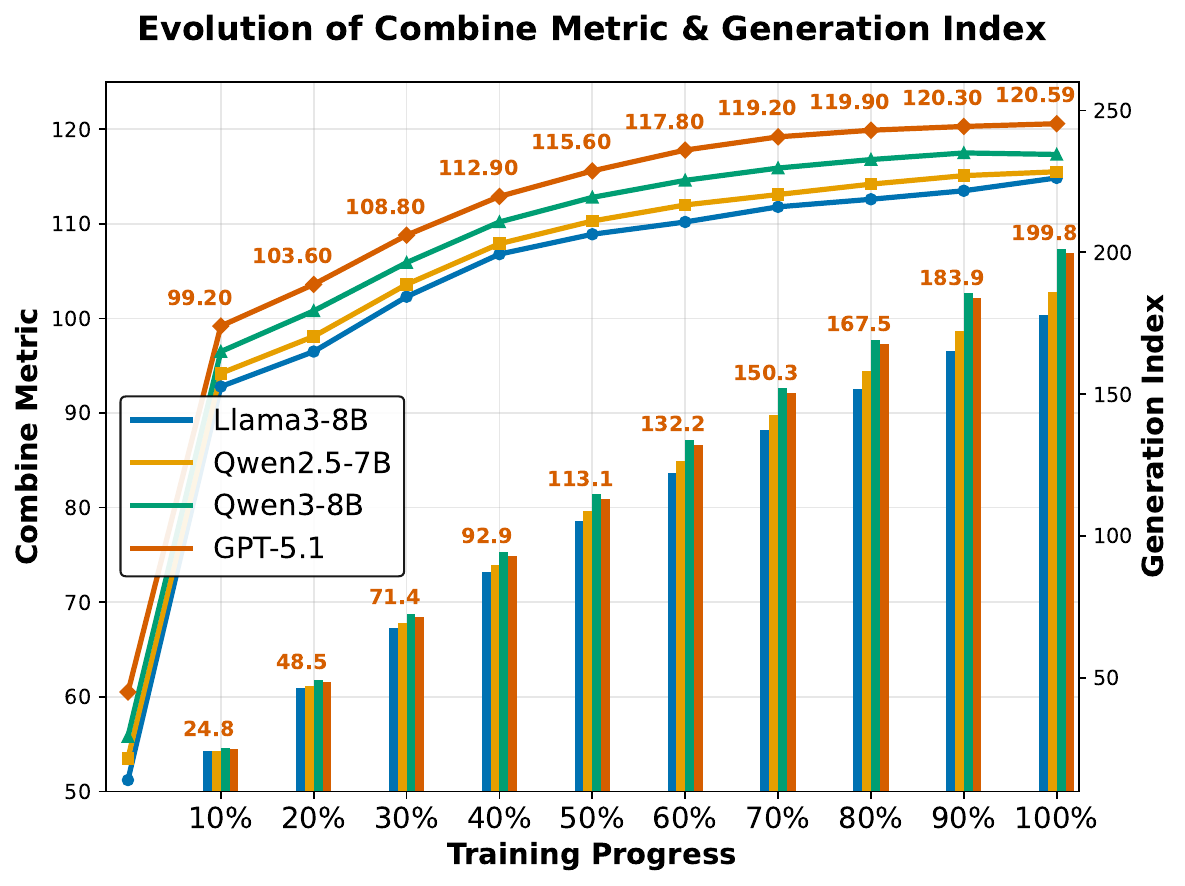}
    \caption{Combine metric evolution across generations on MultiWOZ 2.0. All backbones show monotonic improvement, and the rapid early gains reflect the exploration-exploitation trade-off inherent in evolutionary optimization.}
    \label{fig:comb_generatederation}
\end{figure}

\begin{figure}[t]
    \centering
    \includegraphics[width=1\linewidth]{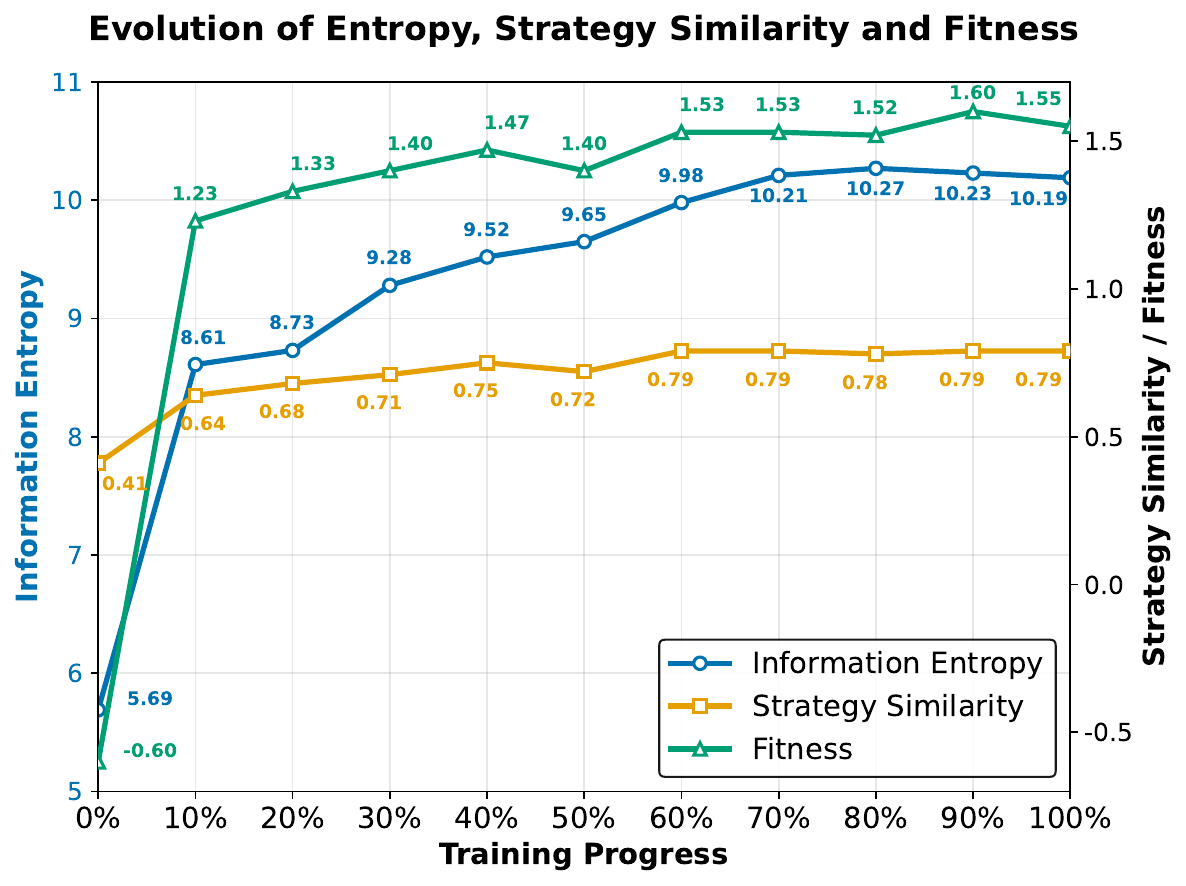}
    \caption{Evolutionary dynamics of ESB across generations on MultiWOZ 2.0 with Qwen3-8B. The simultaneous rise and subsequent decline of entropy and fitness, coupled with increasing pairwise similarity, demonstrates a self organizing transition from exploratory diversity to exploitative convergence.}
    \label{fig:entropy_evolution}
\end{figure}

The experimental results in Table~\ref{tab:main_result} and Appendix Table~\ref{tab:full_result} demonstrate that DarwinTOD achieves SOTA performance across all MultiWOZ versions, , with additional validation on MultiWOZ 2.4 and SGD-X (Appendix~\ref{subsec:recent_datasets}). This consistent superiority is a validation of the theoretical framework in Sec.~\ref{subsec:theoretical_foundation}, effectively amplifying high-fitness strategies and suppressing poor ones. The improved performance demonstrates that the framework effectively capitalizes on the enhanced comprehension and instruction following capabilities of modern LLMs. In particular, even with fewer parameters, Qwen3-4B outperforms both Llama3-8B and Qwen2.5-7B, highlighting the importance of architectural advancements alongside model scale. Furthermore, the improvement in Combine score across generations (Fig.~\ref{fig:comb_generatederation}) empirically validates the convergence property of the evolutionary Markov chain derived in Appendix~\ref{app:convergence}, demonstrating that the stochastic LLM driven mutation operator is successfully filtered by selection and pruning, leading to expected fitness increase.

The generational dynamics in Fig.~\ref{fig:comb_generatederation} reveal that performance improves rapidly in early generations and slows as task success rates rise, reflecting the exploration exploitation trade-off inherent in evolutionary optimization. To dissect the internal evolutionary mechanisms, we conduct detailed experiments (Appendix~\ref{subsec:evolution_dynamics}), then uncover three complementary patterns: (1) DP strategies undergo significantly more revisions than other modules (Table~\ref{tab:generation_index}), indicating evolutionary focus on the core decision making component; (2) Strategies progressively cluster into domain distinct groups in embedding space (Fig.~\ref{fig:tsne_dp_evolution1} and~\ref{fig:tsne_dp_evolution2}), demonstrating structural self organization; and (3) Information entropy and fitness peak then decline while pairwise similarity rises (Fig.~\ref{fig:entropy_evolution}), reflecting a shift from exploration to exploitation. Together, these patterns reveal that DarwinTOD's improvement stems from systematic knowledge consolidation: it autonomously allocates evolutionary effort to the most impactful module, drives semantic specialization through domain feedback, and balances diversity maintenance with selective convergence. This internal self organization, mirrored in the external performance gains, confirms that DarwinTOD achieves genuine lifelong learning through structured evolutionary dynamics rather than incremental patching.

To evaluate the adaptability of DarwinTOD under data scarcity conditions and in the absence of task-specific training data, we conduct comprehensive few-shot experiments (Appendix~\ref{app:fewshot-result}) and zero-shot experiments (Appendix~\ref{app:zerosht_result}). The results demonstrate that the system can effectively leverage its evolutionary operators to rapidly bootstrap and refine strategies from limited interactions. In both settings, it exhibits strong performance and surpasses previous SOTA methods. This confirms that DarwinTOD achieves robust zero-shot generalization through its modular architecture and explicit reasoning pathways. In practical deployment, its performance on unseen domains can be further enhanced through continuous offline evolution.

Furthermore, the domain-incremental continual learning experiment (Appendix~\ref{subsubsec:continual_learning}) demonstrates that DarwinTOD effectively resists catastrophic forgetting. Introducing new domains sequentially does not degrade performance on previously learned domains. This robustness stems from the ESB's domain-isolated strategy pools and fitness-based selection. These results further underscore the framework's capacity for stable, lifelong adaptation in dynamic environments.

\subsection{Ablation Study}
\label{subsec:ablation_study}

\begin{table}[t]
\caption{Ablation study of DarwinTOD components on MultiWOZ 2.0 with Qwen3-8B backbone. $\Delta$ indicates combine change relative to the full system.}
\label{tab:ablation_main}
\centering
\begin{adjustbox}{width=1\columnwidth}
\begin{tabular}{lcccccc}
\toprule
\textbf{Variant} & \textbf{Inform} & \textbf{Succ.} & \textbf{BLEU} & \textbf{Combine} & $\Delta$ \\
\midrule
DarwinTOD (Full) & 98.34 & 92.86 & 21.74 & 117.34 & -- \\
\midrule
\multicolumn{6}{l}{\textit{Online Execution Variants}} \\
\hspace{2mm} w/o Reasoning & 92.53 & 86.93 & 20.95 & 110.68 & $-6.66$ \\
\hspace{2mm} w/o Peer Critique & 82.79 & 77.78 & 18.74 & 99.03 & $-18.31$ \\
\hspace{2mm} w/ E2E Agent & 78.00 & 78.00 & 17.54 & 95.54 & $-21.80$ \\
\midrule
\multicolumn{6}{l}{\textit{Offline Evolution Variants}} \\
\hspace{2mm} w/o Evolution & 75.25 & 70.96 & 19.17 & 92.28 & $-25.06$ \\
\hspace{2mm} w/o Consolidate & 92.71 & 87.11 & 20.99 & 110.91 & $-6.43$ \\
\hspace{2mm} w/o Prune & 91.56 & 86.01 & 20.73 & 109.51 & $-7.83$ \\
\midrule
\multicolumn{6}{l}{\textit{Selection \& Retrieval Variants}} \\
\hspace{2mm} w/ Roulette Wheel & 94.48 & 88.76 & 21.39 & 113.01 & $-4.33$ \\
\hspace{2mm} w/ Random & 93.50 & 87.84 & 21.17 & 111.84 & $-5.50$ \\
\hspace{2mm} w/ $\epsilon$-Greedy & 95.45 & 89.67 & 21.61 & 114.17 & $-3.17$ \\
\bottomrule
\end{tabular}
\end{adjustbox}
\end{table}

The ablation results (Table~\ref{tab:ablation_main}) indicate that each component of DarwinTOD contributes substantially to overall performance.

\noindent \textbf{Online Execution Variants.} Removing agent reasoning (w/o Reasoning) degrades performance, as subsequent agents cannot integrate prior outputs and rationales, so critiques lack substantive grounding. Removing peer critique (w/o Peer Critique) causes a significant drop, since evolution depends only on sparse end of dialog success signals, whereas per-turn critiques provide dense incremental feedback. Replacing the modular pipeline with a monolithic end-to-end agent (w/ \textit{E2E Agent}) causes largest performance drop, validating the POMDP theory: pipeline modules not only prevent cascading error propagation, but also enhance structured state reasoning for targeted strategy evolution.

\noindent \textbf{Offline Evolution Variants.} Disabling the offline evolutionary loop (w/o \textit{Evolution}) corresponds to the zero-shot setup, where each agent uses a single manually designed domain-agnostic static strategy. Although such strategies can achieve strong performance, they lack the ability to adapt to specific domains or learn from interaction failures and negative critiques. Without consolidation (w/o \textit{Consolidate}) or pruning (w/o \textit{Prune}), the strategy bank becomes redundant and inefficient. This confirms that consolidation and pruning are crucial for maintaining a compact and high-performing ESB.

\noindent \textbf{Selection Variants.} Replacing Boltzmann selection with roulette-wheel, uniform random sampling, or $\epsilon$-greedy ($\epsilon=0.1$) yields lower performance, demonstrating that Boltzmann is superior for balancing exploitation of high-fitness strategies with exploration of newer ones. These results confirm that effective lifelong learning requires an explicit balance between exploration and exploitation.

\subsection{Human and Real World Evaluation}

To complement automated metrics, we conduct four complementary evaluations (Appendix~\ref{app:human_studies}) assessing the practical utility, safety, and user experience of DarwinTOD's evolved strategies. An expert evaluation (Fig.~\ref{fig:human_eval_radar}) rated strategies from different LLM backbones in five dimensions. Results show evolved strategies achieve high scores in all backbones, with safety and interpretability being particularly robust, indicating that evolutionary pressure combined with structured peer critique inherently promotes aligned and comprehensible policies.

The real-user study (Table~\ref{tab:real_user_results}) shows the second half of dialogs exhibited a 3.51 point increase in success rate and a reduction of 2.44 turns on average, demonstrating that the system improves continuously through interaction. To further validate this lifelong self-evolution capability in fully autonomous, open-ended settings, we conduct large-scale simulated user experiments (Appendix D.3), where the user simulator autonomously generates complete dialogs based on diverse goals and personalities. As reported in Appendix Table~\ref{tab:simulated_user_results}, DarwinTOD achieves consistent and significant improvements in task success and turn efficiency over hundreds of dialogs, closely mirroring the real-user findings and confirming that the system can evolve without human intervention.

An adversarial and off-topic input test (Table~\ref{tab:robustness_results}) confirms that the system maintains task focus and generates safe, natural responses even under challenging input. These results collectively indicate that DarwinTOD's self-evolution translates into tangible gains in real-world usability and user perception, while its architecture inherently ensures robustness and goal-directed recovery without cascading failures. This work bridges the gap between autonomous optimization and practical deployment, offering a path toward conversational agents that continuously improve through interaction while remaining robust and aligned.

\section{Analysis}

Supplementary experiments (Appendix~\ref{app:supp_exp}) reveal how the dual-loop architecture balances autonomy, robustness, and efficiency. The initialization analysis (Table~\ref{tab:supp_initial}) confirms that the system is robust to various starting conditions, whether minimal descriptions, small strategy banks, or even expert curated seeds. Performance converges to near expert levels regardless of initial quality, demonstrating that the evolutionary loop effectively compensates for sparse prior knowledge. This aligns with the theoretical Markov chain model (Appendix~\ref{app:convergence}): fitness driven selection and pruning gradually filter out noisy mutations, enabling the population to self organize toward high performance strategies without manual tuning. The result underscores a fundamental advantage of evolutionary learning over fixed or episodic update paradigms: the system can bootstrap and refine dialog policies from near-zero knowledge, reducing dependency on costly human curation.

The sensitivity study of Boltzmann temperature $\tau$ (Fig.~\ref{fig:tau_sensitivity}) highlights the trade-off between exploration and exploitation. Optimal performance at $\tau = 1.0$ reflects balance, while excessively low $\tau$ leads to premature convergence, and high $\tau$ slows progress by diluting selection pressure. This mirrors classic evolutionary algorithms and reinforcement learning, yet here the balance is managed implicitly through a single interpretable parameter. The finding validates that DarwinTOD's selection mechanism not only supports sustained improvement but also offers a tunable knob for adapting to different environments, a property essential for deployment in dynamic real-world settings.

The cross-model evolution and efficiency analysis (Appendix~\ref{app:system_extension}) highlight the practical scalability and architectural flexibility of DarwinTOD. Using a powerful LLM only for offline evolution can substantially boost performance even when online agents are lightweight, confirming that evolution can be decoupled from execution to optimize both cost and latency. In contrast, aggressive real-time feedback mechanisms such as online arbitration or per-turn evolution offer diminishing returns relative to their computational overhead, which reinforces our choice of post-dialog offline evolution as a sustainable compromise. Together, these results show that the modular dual-loop design of DarwinTOD not only enables autonomous self-improvement but also supports efficient deployment strategies. Specifically, lightweight models handle real-time dialog for low latency, while more capable models perform asynchronous offline evolution. This arrangement provides a practical balance among latency, cost, and performance for production deployment.

To illustrate the internal evolutionary dynamics and error resilience of DarwinTOD, we conduct two case studies in Appendix~\ref{app:case_study}: one traces the evolutionary trajectory of a DP strategy across generations, and another examines how peer critique both prevents cascading errors and catalyzes targeted strategy evolution in a multi-domain dialog. These studies collectively validate that the dual-loop architecture enables continuous strategy refinement and robust error containment through structured evolution and inter-agent critique.

\section{Conclusion}
DarwinTOD advances self-evolving dialog systems by formulating strategy optimization as the population-based evolutionary process. Unlike traditional prompt engineering, which seeks a single best prompt for a static task, our system maintains a diverse and competing set of strategies that continuously adapt through interaction. This shift from point wise prompt tuning to population wise strategy evolution enables truly lifelong and autonomous improvement in dynamic TOD settings. DarwinTOD achieves SOTA performance across benchmarks and demonstrates sustained improvement via structured evolutionary dynamics. Our experiments show that the Evolvable Strategy Bank self organizes into domain specialized and high-fitness strategies, and human evaluations confirm the safety, interpretability, and practical effectiveness of the evolved strategies. This work establishes a scalable foundation for building continuously learning conversational agents, opening avenues for more efficient evolutionary operators and human-in-the-loop adaptation in autonomous interactive systems.

\section*{Limitations}

Although DarwinTOD has been rigorously validated on established TOD benchmarks such as MultiWOZ and SGD, its practical utility may be constrained by the current reliance on simulated interactions. To fully realize task completion in real-world deployments, the framework would benefit from enhanced capabilities in function calling and agent oriented orchestration, enabling robust integration with external systems and dynamic tool execution. Future work should strengthen these aspects to bridge the gap between benchmark performance and operational effectiveness in open environments.

DarwinTOD's dual-loop architecture introduces additional system complexity compared to static end-to-end models. The online multi-agent pipeline with per-turn peer critique increases inference latency proportionally to the number of agents, which may challenge real-time requirements in latency-sensitive applications (e.g., voice assistants). The offline evolutionary loop, while asynchronous, still requires computational resources for strategy mutation, consolidation, and pruning. In resource-constrained or high-frequency interaction environments, the overhead of maintaining an Evolvable Strategy Bank and periodically updating it may become non-negligible. Furthermore, the framework's reliance on LLM APIs or locally hosted models introduces variable latency and cost trade-offs that depend on deployment scale. These factors do not undermine the theoretical contribution of lifelong self-evolution, but they delineate the practical regimes—such as batch processing, non-interactive optimization, or scenarios with relaxed latency budgets—where DarwinTOD is most directly applicable.


\section*{Ethical Considerations}

In this work, we utilized the publicly and already de-identified MultiWOZ and SGD datasets for evaluation. These datasets do not contain personally identifiable information or offensive content, and their use complies with the consent agreements established during their original release. As our study did not involve new data collection from human participants, no ethics review board approval was required. Throughout our extensive experimental evaluations, including human assessments, we did not observe the emergence or amplification of social biases, harmful content, or manipulative behaviors in the evolved strategies. The evolutionary process, governed by structured peer critique and fitness-based selection oriented towards task success, demonstrated an inherent tendency to converge towards effective and neutral dialogue strategies. However, as with any autonomous learning system deploying LLMs, ongoing monitoring in real-world applications remains a recommended practice to safeguard against unforeseen edge cases.

We acknowledge the use of Writeful integrated with Overleaf for refining the textual expression of this manuscript, and DeepSeek V3.2 for error correction of the experimental code. The role of these LLMs was limited to technical assistance and did not involve research ideation or the creation of core content. All LLM outputs have been rigorously verified by the authors, who bear full responsibility for the final accuracy, integrity, and originality of the content including the avoidance of plagiarism or scientific misconduct.

\section*{Acknowledgments}
This work was supported by the Shenzhen Medical Research Fund (No. D2404001), and in part by the National Natural Science Foundation of China (No. 62472277), the Shanghai East Talents Program (2023-177), the Key Research and Development Program of Guangdong Province (No. 2025B1111020001), the Shenzhen Municipal STIB Key programs (No. CJGJZD20230724093303007 and KJZD20240903101259001), the National Key Laboratory of the CAS on Medical Imaging Science and Technology System, the Xisike Clinical Oncology Research Foundation (Y-2024AZ(NSCLC)MS-0156), and the SIAT-WUXI Joint Innov-Group for AGI-MET.


\bibliography{custom}

\appendix

\section*{Appendix Contents}
\startcontents[appendix]
\printcontents[appendix]{l}{1}{\setcounter{tocdepth}{2}}

\section{Theoretical Analysis of Evolutionary Dynamics}
\label{app:complete_theorem}

This appendix provides a formal analysis of the evolutionary dynamics underlying DarwinTOD. We explain how its dual-loop architecture enables continuous improvement despite potential biases in LLM generated critiques and the inherent stochasticity of LLM driven mutation.

\subsection{Foundations of the Evolutionary Process}
\label{app:foundations}

We frame the lifelong learning of ESB as a Markov process across generations. The ESB at generation $t$, denoted $\Pi_t$, evolves to $\Pi_{t+1}$ through selection, feedback accumulation, and evolutionary operators. Each strategy $\pi \in \Pi_t$ is associated with a fitness score $\phi(\pi)$ defined in Eq.~\ref{eq:fitness}, which balances cumulative positive feedback $H_{\pi}^{+}$ against negative feedback $H_{\pi}^{-}$, normalized by usage count $N_{\pi}$ to ensure fair comparison. The term $\alpha \cdot \text{norm}(\pi_{\text{gen}})$ introduces an age penalty that prevents stagnation and encourages exploration of newer variants.

During online execution, a strategy $\pi_i$ applicable to domain $d$ is selected probabilistically according to the Boltzmann distribution (Eq.~\ref{eq:Boltzmann_selection})~\citep{10912015}. The temperature $\tau$ controls the exploration exploitation trade-off: a lower $\tau$ amplifies the selection pressure toward high-fitness strategies, while a higher $\tau$ allows more exploration. This probabilistic selection aligns with fitness proportionate selection in evolutionary computation~\citep{10.5555/240028} and ensures that the population does not prematurely converge to sub-optimal peaks.

\subsection{Robustness to Noisy Critiques}
\label{app:robustness_critique}

A central concern is that the peer critiques generated by LLM agents may contain noise, bias, or self reinforcing errors. DarwinTOD does not rely on the accuracy of any single critique. Instead, it leverages long term statistics at the strategy level and population level selection to filter out such noise.

Formally, let $c_{\pi,t}$ be the critique signal for strategy $\pi$ at turn $t$, which can be decomposed as
\[
c_{\pi,t} = \bar{c}_{\pi} + \eta_{\pi,t},
\]
where $\bar{c}_{\pi}$ is the underlying true quality signal and $\eta_{\pi,t}$ is a noise term that may exhibit bias. The fitness function (Eq.~\ref{eq:fitness}) aggregates critiques over multiple interactions. By the law of large numbers, as $N_\pi$ grows, the sample averages converge to the true expected value up to a bias term:
\[
\lim_{N_\pi \to \infty} \frac{H_\pi^{+}-H_\pi^{-}}{N_\pi} = \mathbb{E}[\bar{c}_{\pi}] + \mathcal{O}\big(\frac{\sigma_{\eta}}{\sqrt{N_\pi}}\big) + \text{bias}(\eta),
\]
where $\sigma_{\eta}$ denotes the variability of the noise and $\text{bias}(\eta)$ captures any systematic deviation. The evolutionary loop only requires that the signal to noise ratio be sufficient for the selection and pruning mechanisms to distinguish better strategies over time.

Boltzmann selection does not use individual $c_{\pi,t}$, but uses the fitness $\phi(\pi)$ to determine selection probabilities. Strategies that receive spuriously high or low critiques in a few dialogs are not permanently advantaged or disadvantaged because the fitness score smooths out short term fluctuations. The periodic pruning operation further removes strategies whose fitness remains low over many generations, providing an additional correction at the population level. Consequently, even if critiques are occasionally biased or erroneous, the combined effect of fitness aggregation, probabilistic selection, and pruning ensures that the ESB evolves toward genuinely better strategies.

\subsection{Convergence Analysis}
\label{app:convergence}

We now analyze the convergence behavior of the ESB under the joint influence of noisy critiques and stochastic mutations. The mutation operator, powered by an LLM, does not guaranty improvement and may produce strategies that are better, worse, or neutral relative to their parents \citep{10.1145/3638530.3654238, 10.1145/3712256.3726450}. Let $\pi'$ be a mutant of $\pi$. The change in fitness can be written as
$$
\Delta \phi = \phi(\pi') - \phi(\pi) = \delta + \xi,
$$
where $\delta$ is the systematic improvement or degradation introduced by the mutation, and $\xi$ is a random variable representing the noise of the mutation.

The evolutionary loop does not require every mutation to be beneficial. Mutations are triggered only when a strategy is involved in a failed dialog or receives negative critiques (Section~\ref{sec:algorithmic_framework}), ensuring that evolutionary effort focuses on identified weaknesses. The consolidation operator merges strategies only when their semantic similarity exceeds threshold $\delta$, producing a consolidated strategy whose fitness is at least the average of its constituents. The pruning operator directly removes the lowest fitness strategies, providing a strict lower bound on population fitness.

To characterize the convergence dynamics, we analyze the expected average fitness at generation $t$: 
$$
\bar{\phi}(\Pi_t) = |\Pi_t|^{-1}\sum_{\pi \in \Pi_t} \phi(\pi).
$$
Let $p_t$ denote the proportion of strategies involved in failed dialogs or receiving negative critiques at generation $t$, and let $\mu_t = \mathbb{E}[\delta]$ be the expected systematic improvement of mutations conditioned on their occurrence. Experimentally, we observe that $p_t$ decreases over generations as ESB accumulates higher-quality strategies (Figure~\ref{fig:entropy_evolution}), while $\mu_t$ becomes increasingly positive as evolutionary pressure refines the mutation operator's effectiveness.

Under these observable conditions, the expected change in average fitness satisfies:
$$
\mathbb{E}[\bar{\phi}(\Pi_{t+1}) \mid \Pi_t] \geq \bar{\phi}(\Pi_t) + p_t \cdot \mu_t - \eta_t,
$$
where $\eta_t$ represents the noise term arising from stochastic mutations and critiques, which diminishes as the population stabilizes. The inequality holds because: (1) selection preferentially replicates high-fitness strategies, (2) consolidation preserves or improves fitness, (3) pruning eliminates only the lowest fitness individuals, and (4) mutations are applied to underperforming strategies with expected improvement $\mu_t$.

As evolution progresses, $p_t$ decreases as fewer strategies need evolve, while $\mu_t$ increases as mutations become more targeted, causing the product $p_t \cdot \mu_t$ to eventually dominate the noise term $\eta_t$. Consequently, the sequence $\{\mathbb{E}[\bar{\phi}(\Pi_t)]\}_{t=0}^{\infty}$ exhibits an overall upward trend, as empirically validated in Figure~\ref{fig:entropy_evolution}, where the average fitness rises despite non-monotonic fluctuations due to exploration. The bounded strategy space and finite population size ensure that this sequence converges to a stable distribution concentrated on high-performing strategies.

\section{Experiment Details}

\subsection{Datasets Description}
\label{app:datasets_details}

This work employs the MultiWOZ and the SGD datasets to evaluate dialog state tracking models. Both datasets are widely recognized and serve as standard benchmarks in the TOD community, allowing direct comparison with established baselines. While MultiWOZ provides a corpus of human-human conversations with multiple refined versions focusing on annotation quality and consistency, SGD offers a larger-scale, synthetically generated dataset designed to test scalability and zero-shot generalization across a diverse and dynamic set of services. The complementary nature of these datasets enables a comprehensive evaluation of the performance of the model under different conditions.

\subsubsection{MultiWOZ}

MultiWOZ 2.0~\citep{budzianowski-etal-2018-multiwoz} is a large-scale, multi-domain, fully-annotated corpus of human-human task-oriented dialogs. It was collected using a Wizard-of-Oz framework and spans seven distinct domains: Restaurant, Hotel, Attraction, Taxi, Train, Hospital, and Police. With more than 10,000 dialogs, it is one of the largest publicly available datasets of its kind and has been widely adopted as a standard benchmark for dialog state tracking (DST) and related tasks. Its scale, multi-domain nature, and rich annotations have made it a foundational resource in the dialog research community.

\begin{table}[t]
\centering
\caption{Key statistics of the MultiWOZ 2.0 dataset.}
\label{tab:multiwoz20_stats}
\begin{tabular}{l r}
\toprule
\textbf{Statistic} & \textbf{Value} \\
\midrule
Total Dialogs & 10,438 \\
Train Dialogs & 8,438 \\
Dev Dialogs & 1,000 \\
Test Dialogs & 1,000 \\
Total Turns & 115,434 \\
Domains & 7 \\
Avg. Turns per Dialog & 13.68 \\
Slots & 25 \\
Slot Values in Ontology & 4,510 \\
\bottomrule
\end{tabular}
\end{table}

\begin{table}[t]
\centering
\caption{Key statistics of the SGD dataset.}
\label{tab:sgd_stats}
\begin{tabular}{l r}
\toprule
\textbf{Statistic} & \textbf{Value} \\
\midrule
Total Dialogs & 22,825 \\
Train Dialogs & 16,142 \\
Dev Dialogs & 2,482 \\
Test Dialogs & 4,201 \\
Total Turns & 467,916 \\
Domains & 16 \\
Services & 45 \\
Avg. Turns per Dialog & 20.44 \\
Slots & 214 \\
Slot Values & 14,139 \\
\bottomrule
\end{tabular}
\end{table}

Following the release of MultiWOZ 2.0, two subsequent versions were introduced to address annotation noise and inconsistencies. MultiWOZ 2.1~\citep{eric-etal-2020-multiwoz} corrected substantial noise in dialog state annotations and user utterances, affecting 32\% of state annotations across 40\% of turns; also canonicalized slot values, added user dialog acts, and included natural language descriptions for each slot to facilitate low-resource and zero-shot learning.

Building upon 2.1, MultiWOZ 2.2~\citep{zang-etal-2020-multiwoz} introduced further refinements: it corrected additional state errors in 17.3\% of utterances, redefined the ontology by splitting slots into categorical and non-categorical types, added slot span annotations for non-categorical slots to support span-based models, and introduced annotations for active user intents and requested slots per-turn. These revisions aim to provide a cleaner, more consistent, and more richly annotated benchmark for robust evaluation of dialog state tracking models.

\subsubsection{Schema-Guided Dialog (SGD)}
The Schema-Guided Dialog (SGD) dataset~\citep{Rastogi_Zang_Sunkara_Gupta_Khaitan_2020} is a large-scale, multi-domain, task-oriented dialog corpus designed to address the scalability challenges of modern virtual assistants. Unlike previous datasets that assume a single static ontology per domain, SGD introduces a schema-guided paradigm where each service provides its own schema containing intents, slots, and natural language descriptions. This approach enables models to handle heterogeneous APIs and facilitates zero-shot generalization to unseen services. The dataset spans 16 distinct domains with 45 distinct services and includes multi-domain conversations that reflect realistic user interactions. With over 16,000 dialogs, SGD is the largest publicly available task-oriented dialog dataset and serves as a benchmark for intent prediction, slot filling, dialog state tracking, and language generation in scalable, multi-service environments.

\subsection{Baselines Details}
\label{app:baselines_details}

To comprehensively evaluate the performance of DarwinTOD, we compare it with a range of representative and state-of-the-art models in TOD systems. These baselines are categorized according to their architectural paradigms, covering early generative models, pre-trained dialog models, and recent LLM-based agent systems.

\subsubsection{Early Generative Models}
These early approaches unify the dialog process into a single sequence-to-sequence or auto-regressive model, based on pre-trained language models, reducing modular dependencies and demonstrating the potential of generative architectures for TOD.
    
\noindent \textbf{SimpleTOD}~\citep{10.5555/3495724.3497418} formulates all TOD sub-tasks as a single causal language modeling problem, generating belief states, system acts, and responses autoregressively within a fixed extract-then-query turn structure.

\noindent \textbf{MinTL}~\citep{lin-etal-2020-mintl} employs a transfer learning framework with Levenshtein belief spans to jointly learn DST and response generation in an end-to-end manner.

\noindent \textbf{SOLOIST}~\citep{peng-etal-2021-soloist} integrates DST and response generation into a single auto-regressive language model, though it retains a turn-by-turn state extraction and database query paradigm.
    
\noindent \textbf{UBAR}~\citep{Yang_Li_Quan_2021} fine-tunes GPT-2 on complete dialog sessions, including belief states, database results, and system acts-to achieve fully end-to-end dialog modeling.

\subsubsection{Pre-trained TOD Models}
These models leverage pre-trained language models on large scale dialog corpora to acquire general dialog abilities before task-specific fine-tuning. They represent the dominant paradigm before the rise of large scale instruction-tuned LLMs.

\noindent \textbf{BORT}~\citep{sun-etal-2022-bort} introduces back and denoising reconstruction strategies to improve dialog state accuracy and robustness against error propagation. It employs a T5-small backbone and achieves strong performance particularly in low-resource and zero-shot settings.

\noindent \textbf{PPTOD}~\citep{su-etal-2022-multi} is a unified plug-and-play model based on T5, pre-trained on multiple dialog tasks (DST, DPL, NLG) using a multi-task prompt setup.
    
\noindent \textbf{GALAXY}~\citep{he2022galaxy} is a generative pre-trained model that explicitly injects dialog policy learning through semi-supervised training on both labeled and unlabeled dialog data.
    
\noindent \textbf{ZS-TOD}~\citep{Mosharrof2023ZeroShotGE} is a zero-shot generalizable end-to-end ToD system that takes advantage of domain schemas and dialog state summarization to enable robust generalization to unseen domains.

\noindent \textbf{TOATOD}~\citep{bang-etal-2023-task} employs task-optimized adapters on T5, enabling efficient adaptation to different TOD tasks with minimal parameter updates.

\noindent \textbf{KMc-ToD}~\citep{10.1016/j.knosys.2024.111662} integrates a multi-copy mechanism with a structured schema graph to enhance slot selection and consistency in end-to-end response generation, improving the integration of domain specific slots into delexicalized responses.

\subsubsection{Recent LLM-based and Agent-Oriented Models}
These models leverage large language models as core controllers or autonomous agents, supporting flexible interaction with external tools and APIs, and representing the latest paradigm shift towards more open and adaptive dialog systems.

\noindent \textbf{AutoTOD}~\citep{xu-etal-2024-rethinking} is a fully zero-shot autonomous agent that abandons traditional modular design, relying solely on an instruction-following LLM guided by a schema to dynamically decide API calls and generate responses.

\noindent \textbf{OmniDialog}~\citep{razzhigaev-etal-2024-omnidialog} is a multimodal pre-trained model that unifies dialog comprehension, management, and generation within a multi-task framework, demonstrating strong low-resource and cross-domain transfer ability.

\noindent \textbf{ProTOD}~\citep{dong-etal-2025-protod} is a proactive TOD framework that uses an adaptive exploratory retrieval mechanism and a two-stage policy planner to dynamically explore domain knowledge and plan multi-task dialogs.
    
\noindent \textbf{AgentTOD}~\citep{10.1145/3745021} employs an LLM as a controller to dynamically decide when and how to call external API, supporting multiple API calls per-turn and adapting to complex user queries without a fixed extract query paradigm.

\subsection{Discussion on Evaluation Metrics}
\label{app:metrics_discussion}

BLEU is an imperfect metric for evaluating natural language generation, particularly for modern LLMs, as it primarily measures n-gram overlap with a single reference and often fails to capture semantic equivalence, fluency, or task appropriateness. This limitation is well-recognized in the broader NLP community.

However, we emphasize the context of our evaluation within the TOD field, where standardized protocols are essential for fair comparison with a large body of prior work. The combined score, defined as $(\text{Inform} + \text{Success}) \times 0.5 + \text{BLEU}$, has been the standard composite metric for end-to-end TOD evaluation for over five years. It has been consistently adopted by nearly all relevant baselines from 2020 to 2025, including SimpleTOD, UBAR, GALAXY, and the previous SOTA AgentTOD. To ensure a fair and reproducible quantitative comparison with these dozens of historical systems, we are obligated to adhere to this widely accepted protocol. Reporting results without BLEU would preclude meaningful comparison with the existing literature.

Further more, our evaluation does not rely solely on automatic metrics. Recognizing the limitations of BLEU, we conducted comprehensive human evaluations to validate the real-world quality of our system. As detailed in Appendix~\ref{app:human_studies}, these studies assessed dimensions that BLEU cannot capture, including interpretability, safety, fluency, and user satisfaction. The results of these human studies strongly corroborate the findings from our automatic metrics.

For the SGD dataset, we report GLEU instead of BLEU, following the convention established by prior zero-shot baselines like ZS-TOD and AgentTOD. GLEU offers a slightly more robust correlation with human judgment by incorporating a recall component, but it shares similar fundamental limitations as BLEU.

\subsection{Detailed Experimental Setup}
\label{app:detailed_setup}

\begin{table}[t]
\caption{Hyperparameter settings for DarwinTOD experiments.}
\label{tab:hyperparameters}
\centering
\begin{adjustbox}{width=1\columnwidth}
\small
\begin{tabular}{l|l}
\hline
\textbf{Variable} & \textbf{Description} \\
\hline
$\alpha = 0.3$ & Weight for generation penalty (Eq.~\ref{eq:fitness}) \\
$\epsilon = 0.01$ & Smoothing constant in fitness (Eq.~\ref{eq:fitness}) \\
$\tau = 1.0$ & Temperature for Boltzmann selection (Eq.~\ref{eq:Boltzmann_selection}) \\
$\delta = 0.8$ & Similarity threshold for strategy consolidation \\
$K = 10$ & Number of initial strategies generated per domain \\
$M = 10$ & Maximum number of strategies allowed per domain \\
$\theta_e = 0.8$ & Sampling temperature for evolutionary operators \\
$\theta_x = 0.7$ & Sampling temperature for online dialog agents \\
$T = 30$ & Maximum number of turns per dialog episode \\
\hline
\end{tabular}
\end{adjustbox}
\end{table}

DarwinTOD begins with an empty ESB. During evaluation, for each previously unseen domain encountered in the dialog sequence, the Genesis operator synthesizes $K=10$ distinct strategies based solely on the domain name and the agent's role description, without any task-specific fine-tuning or in context examples (see Appendix~\ref{app:manual_strategies} for the generic strategies used in the zero-shot setup). When a dialog involves a previously unseen combination of multiple domains, the system first checks whether there are strategies for each constituent domain. If a domain lacks strategies, the Genesis operator is invoked to generate them. Subsequently, a new composite strategy is created by randomly selecting and merging one strategy from each involved domain, ensuring the system can handle multi-domain interactions from the outset.

The online execution proceeds as follows: for each dialog turn, the dialog manager reads the user utterance from the training set, along with the preceding dialog history. Each of the three specialized agents (DST, DP, NLG) retrieves a strategy from the ESB using Boltzmann selection (Eq.~\ref{eq:Boltzmann_selection}) with temperature $\tau=1.0$, executes its function with built-in peer critique, and logs the interaction to SSM. After an entire dialog episode concludes, offline evolutionary operations (Genesis, Mutation, Consolidation, Pruning) are triggered based on the accumulated interaction feedback. Semantic similarity is computed via cosine similarity between bge-small-en-v1.5 sentence transformer embeddings, with $\delta = 0.8$ as the merging threshold. The Pruning operator maintains a maximum of $M=10$ strategies per domain. Throughout this process, the metadata (usage counts, feedback scores) of each involved strategy is updated accordingly.

We perform a performance evaluation in a phased manner. Before any training begins, the system is evaluated on the complete test set to establish a baseline. Subsequently, as dialogs are processed sequentially from the training set, the system undergoes periodic evaluation: after every 10\% of the training dialogs have been processed, the updated system is evaluated on the full test set. We employ Llama3-8B, Qwen2.5-7B, Qwen3-8B, and GPT-5.1 as backbone LLMs for both online execution and offline evolution. During online execution, we set the sampling temperature at $\theta_x=0.7$ for agent responses; for evolutionary operators, we use $\theta_e=0.8$. All other hyperparameters are listed in Table~\ref{tab:hyperparameters}.

The system is implemented in Python 3.10, with all open-source LLMs running locally on NVIDIA A100 80 GB, while GPT models are accessed via the OpenAI API.

\section{Detailed Main Experimental Results}
\label{app:detailed_results}

\begin{table*}[t!]
\caption{Performance evaluation and scaling analysis of DarwinTOD across various LLM backbones on MultiWOZ 2.0, 2.1, and 2.2, compared against previous SOTA AgentTOD.}
\label{tab:full_result}
\centering
\begin{adjustbox}{max width=\textwidth}
\begin{tabular}{l|cccc|cccc|cccc}
\hline
\multirow{2}{*}{Model} & \multicolumn{4}{c|}{MultiWOZ 2.0} & \multicolumn{4}{c|}{MultiWOZ 2.1} & \multicolumn{4}{c}{MultiWOZ 2.2} \\
\cline{2-13}
& Inform & Succ. & BLEU & Comb. & Inform & Succ. & BLEU & Comb. & Inform & Succ. & BLEU & Comb. \\
\hline
AgentTOD (Llama3-8B) & 94.84 & 87.21 & 21.97 & 113.00 & 96.89 & 89.38 & 19.91 & 113.04 & 90.71 & 81.26 & 18.02 & 104.00 \\
\hline
DarwinTOD (Llama3-8B) & 96.92 & 89.14 & 21.83 & 114.86 & 98.73 & 91.42 & 19.96 & 115.04 & 92.58 & 83.97 & 17.98 & 106.26 \\
DarwinTOD (Llama3.1-8B) & 96.57 & 90.19 & 21.69 & 115.07 & 97.97 & 91.51 & 20.32 & 115.06 & 92.59 & 84.09 & 18.27 & 106.61 \\
DarwinTOD (Qwen2.5-3B) & 90.42 & 82.17 & 18.96 & 105.26 & 92.15 & 83.84 & 18.42 & 106.42 & 86.31 & 77.95 & 17.21 & 99.34 \\
DarwinTOD (Qwen2.5-7B) & 97.63 & 90.28 & 21.55 & 115.51 & 98.92 & 91.85 & 20.18 & 115.57 & 92.14 & 84.33 & 18.34 & 106.58 \\
DarwinTOD (Qwen2.5-14B) & 97.85 & 91.07 & 21.89 & 116.35 & 99.18 & 92.64 & 20.67 & 116.58 & 93.02 & 85.41 & 18.76 & 107.98 \\
DarwinTOD (Qwen3-4B) & 97.21 & 90.14 & 21.23 & 114.91 & 98.76 & 91.62 & 20.04 & 115.23 & 91.87 & 84.08 & 18.12 & 106.10 \\
DarwinTOD (Qwen3-8B) & 98.34 & 92.86 & 21.74 & 117.34 & 99.62 & 94.18 & 20.33 & 117.23 & 94.73 & 87.25 & 18.41 & 109.40 \\
DarwinTOD (Qwen3-14B) & 98.67 & 93.45 & 22.08 & 118.14 & 99.84 & 94.92 & 20.88 & 118.26 & 95.21 & 87.98 & 18.92 & 110.52 \\
DarwinTOD (GPT-4) & 99.05 & 95.87 & 22.93 & 120.39 & \textbf{99.76} & 96.38 & 21.46 & 119.53 & 96.14 & 89.63 & 21.87 & 114.76 \\
DarwinTOD (GPT-5.1) & \textbf{99.10} & \textbf{96.20} & \textbf{22.94} & \textbf{120.59} & 99.40 & \textbf{96.50} & \textbf{22.19} & \textbf{120.14} & \textbf{96.48} & \textbf{90.12} & \textbf{21.98} & \textbf{115.28} \\
\hline
\end{tabular}
\end{adjustbox}
\end{table*}

\begin{table*}[t]
\centering
\caption{Few-shot evaluation results of DarwinTOD and baseline models on MultiWOZ 2.0 with varying amounts of training data (5\%, 10\%, 20\%)}
\label{tab:multiwoz_fs_results}
\begin{adjustbox}{max width=\textwidth}
\begin{tabular}{l|cccc|cccc|cccc}
\hline
\multirow{2}{*}{Model} & \multicolumn{4}{c|}{5\% Train Data} & \multicolumn{4}{c|}{10\% Train Data} & \multicolumn{4}{c}{20\% Train Data} \\
\cline{2-13}
& Inform & Succ. & BLEU & Comb. & Inform & Succ. & BLEU & Comb. & Inform & Succ. & BLEU & Comb. \\
\hline
SOLOIST & 69.30 & 52.30 & 11.80 & 72.60 & 69.90 & 51.90 & 14.60 & 70.50 & 74.00 & 60.10 & 15.24 & 82.29 \\
MinTL & 75.48 & 60.96 & 13.98 & 82.19 & 78.08 & 66.87 & 15.46 & 87.94 & 82.48 & 68.57 & 13.00 & 88.52 \\
PPTOD & 79.86 & 63.48 & 14.89 & 86.56 & 84.42 & 68.36 & 15.57 & 91.96 & 84.94 & 71.70 & 17.01 & 95.33 \\
UBAR & 73.04 & 60.28 & 16.03 & 82.69 & 79.20 & 68.70 & 16.09 & 90.04 & 82.50 & 66.60 & 17.72 & 92.07 \\
GALAXY & 80.59 & 67.43 & 17.39 & 91.40 & 87.00 & 75.00 & 17.65 & 98.65 & 89.55 & 75.85 & 17.54 & 100.24 \\
AgentTOD (Llama3-8B) & 83.18 & 68.89 & 18.86 & 95.41 & 89.54 & 77.81 & 19.27 & 102.95 & 91.10 & 77.55 & 19.51 & 103.83 \\
\hline
DarwinTOD (Llama3-8B) & 85.26 & 70.82 & \textbf{18.84} & 96.88 & 91.62 & 79.74 & \textbf{19.31} & 104.99 & 93.18 & 79.48 & \textbf{19.51} & 105.84 \\
DarwinTOD (Qwen2.5-7B) & 85.89 & 71.73 & 18.60 & 97.41 & 92.29 & 80.78 & 19.06 & 105.60 & 93.86 & 80.50 & 19.26 & 106.44 \\
DarwinTOD (Qwen3-8B) & \textbf{86.51} & \textbf{73.62} & 18.76 & \textbf{98.83} & \textbf{92.97} & \textbf{82.89} & 19.23 & \textbf{107.16} & \textbf{94.55} & \textbf{82.62} & 19.43 & \textbf{108.02} \\
\hline
\end{tabular}
\end{adjustbox}
\label{tab:fewshot_multiwoz}
\end{table*}

The main experiments include comprehensive evaluations on both MultiWOZ and SGD datasets, covering both few-shot and zero-shot settings. All baseline results are taken from their original papers.

\subsection{Complete Performance Results on MultiWOZ}
\label{app:full_multiwoz}

We provide an extended performance analysis that includes a wider range of backbone LLMs, with Llama3, Llama3.1, Qwen2.5, and Qwen3 for their extended contexts and enhanced abilities. The results in Table~\ref{tab:full_result} allow for a finer grained examination of how model scale, architecture, and evolutionary training interact, offering additional insights into the robustness and scalability of DarwinTOD beyond the condensed summary in the main text. These comprehensive results further substantiate our claim that the proposed framework achieves strong and consistent performance gains across diverse model capacities.

\subsection{Evolution Dynamics Analysis}
\label{subsec:evolution_dynamics}

This subsection aims to examine the intrinsic evolution of ESB, thereby validating the theoretical analysis presented in Section~\ref{subsec:theoretical_foundation} and Appendix~\ref{app:complete_theorem}. To this end, we conduct three complementary analyzes: quantifying update frequency across modules, visualizing semantic trajectories in embedding space, and measuring structural entropy evolution.

\subsubsection{Semantic Trajectory of Evolution}
\label{subsubsec:semantic_trajectory}

\begin{figure}[t]
    \centering
    \includegraphics[width=1\columnwidth]{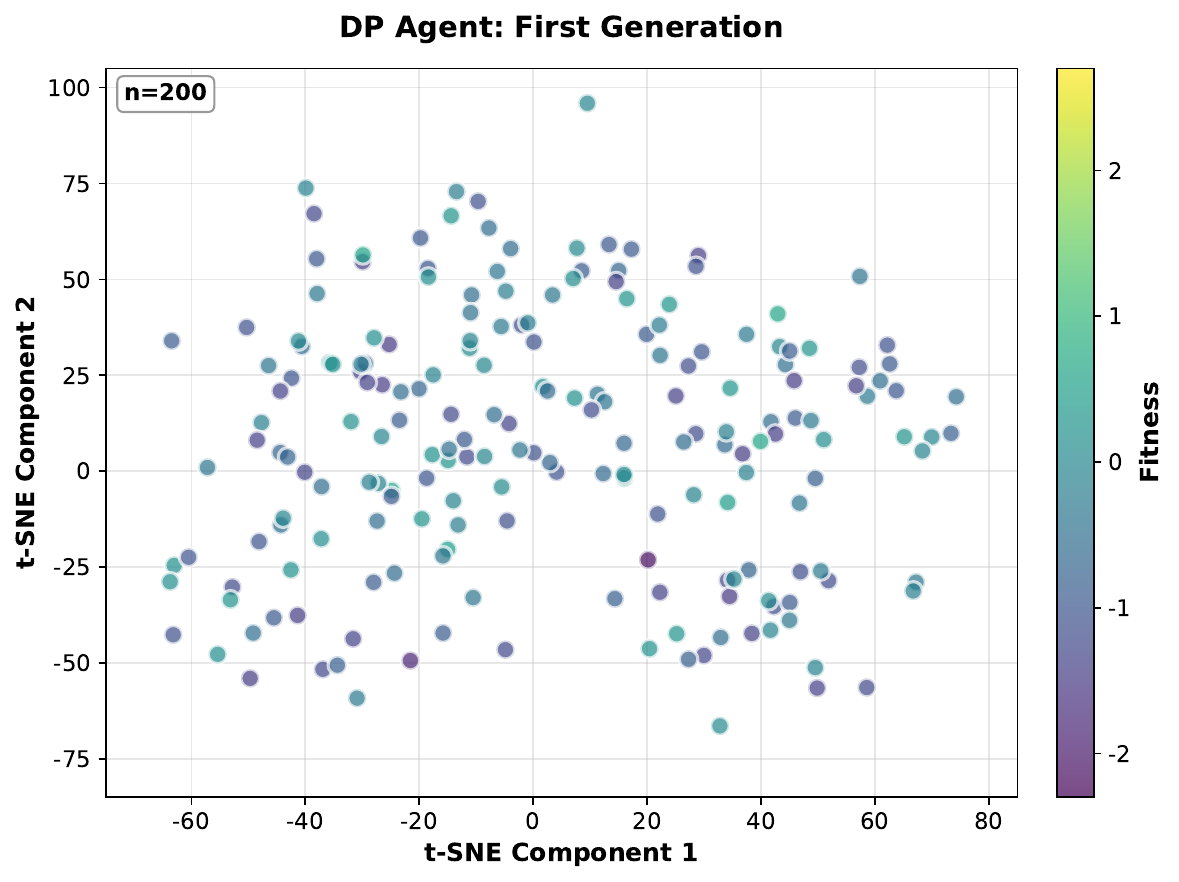}
    \caption{t-SNE visualization of DP strategy embeddings: initial population.}
    \label{fig:tsne_dp_evolution1}
\end{figure}

\begin{figure}[t]
    \centering
    \includegraphics[width=1\columnwidth]{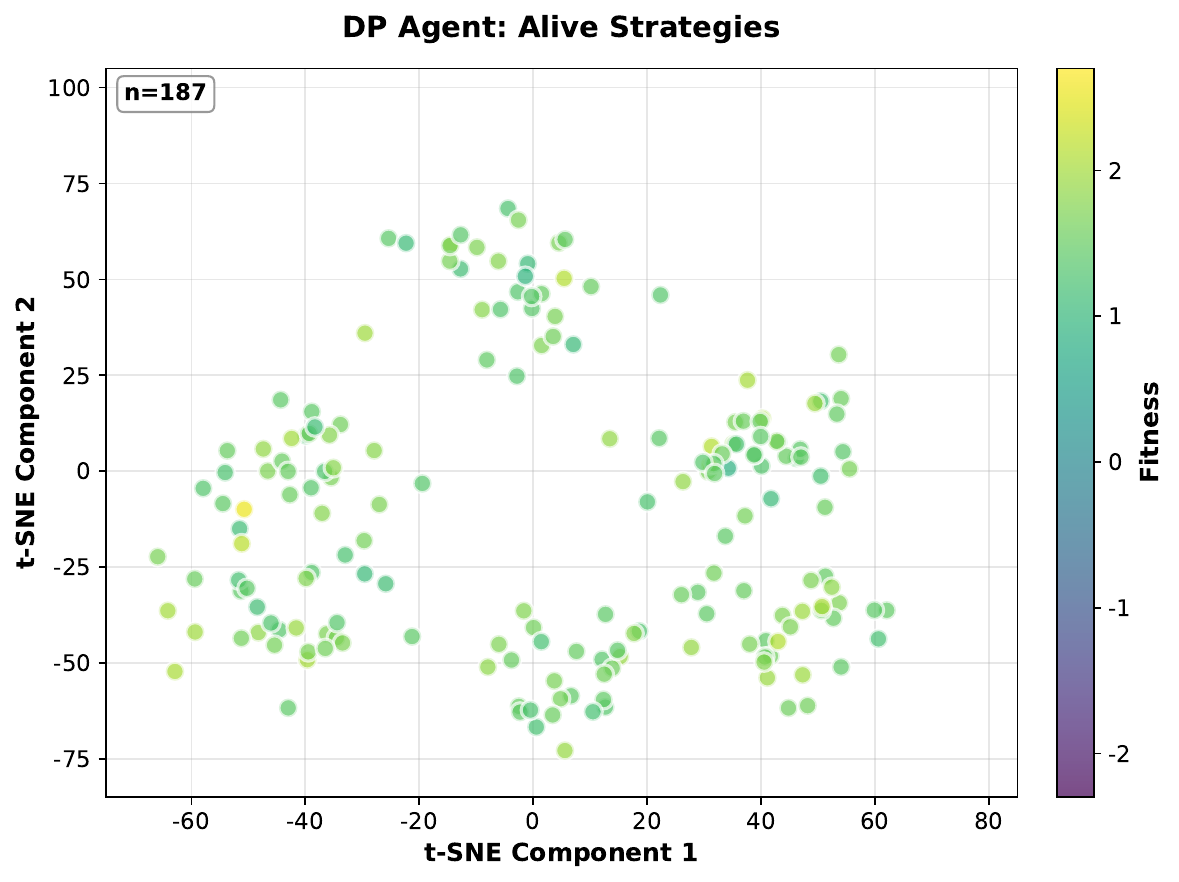}
    \caption{t-SNE visualization of DP strategy embeddings: final alive population.}
    \label{fig:tsne_dp_evolution2}
\end{figure}

To visualize the structural evolution of the strategy bank and examine how the ESB transitions from an initial unstructured state to an organized domain-specialized configuration, we project the sentence embeddings of DP strategies into two dimensions using t-SNE~\citep{vandermaaten2008a}. This dimensionality reduction technique allows us to observe the formation of clusters in semantic space, which correspond to groups of strategies that have evolved to address similar dialog contexts or domain specific challenges. Figure~\ref{fig:tsne_dp_evolution1} and~\ref{fig:tsne_dp_evolution2} contrast the initial (Generation~1) and final alive populations under Qwen3-8B on MultiWOZ 2.0.

The visual comparison reveals a clear evolutionary trajectory. The initial strategies, though generated with domain names via the Genesis operator, are broadly scattered with substantial overlap, reflecting surface level, template like policies that lack deep domain specific adaptation. After evolution, strategies coalesce into several well separated clusters, each corresponding to a distinct domain or coherent multi-domain combination. This structural organization demonstrates that evolutionary pressure, guided by domain specific interaction feedback and consolidation, drives the strategy population to self-organize into highly specialized and efficient policies that are semantically distinct across different domains. The emergence of these domain correlated clusters validates the theoretical convergence properties derived in Appendix~\ref{app:complete_theorem}: the Markov chain evolutionary process, through fitness based selection and semantic merging, progressively concentrates the strategy bank on high-performing, domain adapted policies, thereby materializing the theoretical prediction that the ESB transitions from an initial, weakly differentiated state to a structured, domain optimized configuration.

\subsubsection{Entropy and Structural Evolution}
\label{subsubsec:entropy_analysis}

To quantitatively characterize the structural evolution of the ESB during training, we introduce information entropy as a core analytical metric. Entropy measures the complexity and diversity of strategies; pairwise similarity reflects semantic convergence within the population; and average fitness captures the effectiveness of evolutionary pressure. By simultaneously tracking the dynamics of these three curves, we directly validate the evolutionary principles outlined in Section~\ref{subsec:theoretical_foundation}: the system autonomously balances exploration (maintaining diversity) and exploitation (focusing on high-fitness strategies), ultimately achieving structured convergence through knowledge consolidation.

We quantify the lexical distribution of strategy descriptions in each active strategy bank $\Pi_t$ using Shannon entropy:
$$
H^{(t)} = - \sum_{w \in \mathcal{V}^{(t)}} q(w) \log_2 q(w),
$$
where $q(w)$ is the global unigram frequency distribution throughout the bank, and $\mathcal{V}^{(t)}$ is the unique vocabulary in $\Pi_t$. This entropy value increases with greater lexical diversity or complexity of strategy texts and decreases as descriptions converge, directly reflecting the degree of structural organization in the strategy bank at the content level.

The results (Section~\ref{sec:main_results} Figure~\ref{fig:entropy_evolution}) reveal a clear self-organizing evolutionary path from disordered exploration to ordered convergence. In early generations, entropy and fitness rise simultaneously, corresponding to an exploration phase in which mutation and genesis introduce diverse strategies, expanding the semantic space. As evolution progresses, Boltzmann selection amplifies high-fitness strategies, while consolidation merges semantically similar ones, leading to steadily increasing pairwise similarity and a subsequent decline in both entropy and fitness peaks after reaching a maximum. This dynamic indicates a spontaneous shift from broad exploration to focused exploitation: population diversity is selectively compressed, and knowledge is distilled into a compact set of efficient, refined strategy archetypes. This not only confirms the theoretical analysis in Section~\ref{subsec:theoretical_foundation} that the Markov chain evolution of the ESB progressively converges to high-performance subsets, but also unveils the complete internal evolutionary dynamics of DarwinTOD: evolutionary pressure drives population self organization, transforming initially homogeneous strategy descriptions into a highly specialized structured strategy system. This internal self organization, mirrored by the external performance gains in Table~\ref{tab:main_result} and~\ref{tab:full_result}, jointly demonstrates that DarwinTOD achieves continuous autonomous improvement through structured evolution rather than incremental patching of static strategies.

\subsubsection{Update Frequency across Modules}
\label{subsubsec:update_frequency}

\begin{table}[t]
\centering
\caption{Average generation index of strategies in the final alive population across dialog modules. Higher values indicate more frequent evolutionary updates during the lifespan. Results are on MultiWOZ 2.0.}
\label{tab:generation_index}
\begin{adjustbox}{width=1\columnwidth}
\begin{tabular}{l|cccc}
\hline
\textbf{Backbone LLM} & \textbf{DST Agent} & \textbf{DP Agent} & \textbf{NLG Agent} & \textbf{Avg.} \\
\hline
Llama3-8B       & 62.35 & 159.18 & 8.07   & 76.53 \\
Qwen2.5-7B      & 78.14 & 193.21 & 10.43  & 93.93 \\
Qwen3-8B        & 82.79 & 205.67 & 12.61  & 100.36 \\
GPT-5.1         & 81.06 & 198.32 & 10.15  & 96.51 \\
\hline
\end{tabular}
\end{adjustbox}
\end{table}

This experiment measures how DarwinTOD allocates evolutionary effort across online agents by tracking the average generation index of strategies in the final alive population. Higher values indicate more frequent refinement. The results show that the DP strategies undergo the most intensive updates, aligning with their greater complexity and impact on dialog success.

Table~\ref{tab:generation_index} reveals a consistent pattern: Dialog Policy (DP) strategies are updated significantly more frequently than other agents. It indicates that the evolutionary process self-organizes to concentrate effort on the core decision making module, whose strategies are both more complex to optimize and have a higher leverage on dialog success. The higher update frequency for DP reflects its role in navigating a vast space of multi-turn planning decisions, whereas the relative stability of NLG strategies suggests the system converges on reusable, template like responses that require less domain specific tuning. Furthermore, the positive correlation between overall update activity and backbone LLM capability demonstrates that more capable models not only propose better initial strategies but also engage in more prolific and effective refinement, validating that the framework's evolutionary capacity scales with the underlying LLM's reasoning power.

\subsection{Evaluation for Few-shot Capability}
\label{app:fewshot-result}

We conduct few-shot evaluations to assess DarwinTOD's ability to adapt and perform effectively when only a limited amount of task-specific training data is available. In real-world deployment scenarios, abundant annotated dialogs are often scarce, especially for new or niche domains. Therefore, it is crucial to examine how well a lifelong self-evolving system can bootstrap and improve under low-resource conditions. The experiments summarized in Table~\ref{tab:multiwoz_fs_results} measure performance using only 5\%, 10\% and 20\% of the MultiWOZ 2.0 training data, providing insight into the data efficiency and rapid adaptation capability of the framework in comparison to strong static baselines.

\subsection{Evaluation for Zero-shot Capability}
\label{app:zerosht_result}

\begin{table}[t]
\caption{Zero-shot performance comparison on MultiWOZ 2.0 Dataset.}
\label{tab:multiwoz_zs_results}
\centering
\begin{adjustbox}{width=1\columnwidth}
\begin{tabular}{l|ccccc}
\hline
Model & Inform & Succ. & BLEU & Comb. \\
\hline
GALAXY & 9.10 & 4.76 & 3.61 & 10.54 \\
TOATOD & 10.59 & 5.21 & 3.83 & 11.73 \\
AgentTOD (Llama3-8B) & 67.50 & 45.98 & 14.24 & 70.98 \\
\hline
DarwinTOD (Llama3-8B) & 74.18 & 68.26 & \textbf{19.25} & 90.47 \\
DarwinTOD (Qwen2.5-7B) & 74.72 & 69.12 & 19.00 & 90.92 \\
DarwinTOD (Qwen3-8B) & \textbf{75.25} & \textbf{70.96} & 19.17 & \textbf{92.28} \\
\hline
\end{tabular}
\end{adjustbox}
\end{table}

\begin{table}[t]
\caption{Zero-shot performance comparison on SGD Dataset.}
\label{tab:sgd_zs_results}
\centering
\begin{adjustbox}{width=1\columnwidth}
\begin{tabular}{l|cccc}
\hline
Model & Inform & Succ. & GLEU & Comb. \\
\hline
SimpleTOD & 68.10 & 60.48 & 20.40 & 84.69 \\
ZS-TOD & 71.68 & 61.63 & 18.51 & 85.16 \\
AgentTOD (Llama3-8B) & 76.70 & 65.82 & \textbf{21.66} & 92.92 \\
\hline
DarwinTOD (Llama3-8B) & 80.12 & 71.25 & 19.11 & 94.80 \\
DarwinTOD (Qwen2.5-7B) & 80.71 & 72.16 & 18.86 & 95.30 \\
DarwinTOD (Qwen3-8B) & \textbf{81.29} & \textbf{74.06} & 19.03 & \textbf{96.71} \\
\hline
\end{tabular}
\end{adjustbox}
\end{table}

We conduct zero-shot evaluations on both MultiWOZ (Table~\ref{tab:multiwoz_zs_results}) and SGD (Table~\ref{tab:sgd_zs_results}) datasets to rigorously assess DarwinTOD's baseline performance without any in-domain training data. In this setting, the system starts with an empty ESB. For execution, each agent is initialized with a single domain-agnostic strategy that was manually designed prior to evaluation (see the Appendix~\ref{app:manual_strategies}). No strategy selection or evolutionary refinement is performed; each agent uses the same static strategy throughout all dialogs. We employ the GLEU metric~\citep{wang-etal-2018-glue} in the SGD data set, consistent with the three baselines for a fair comparison. GLEU represents an improved version of the BLEU~\citep{papineni-etal-2002-bleu} metric and offers a more comprehensive assessment of language similarity.

\subsection{Evaluation on More Recent Datasets}
\label{subsec:recent_datasets}

To further validate the generalization capability of DarwinTOD beyond the commonly used benchmarks, we additionally conduct experiments on two more recent datasets released in 2022. These datasets are designed to test model performance under cleaner annotations and linguistic variations. Since no baseline models have reported results on these datasets, we present the performance of DarwinTOD with different backbones as standalone references.

\subsubsection{MultiWOZ 2.4 (2022)}
MultiWOZ 2.4~\citep{ye-etal-2022-multiwoz} is a refined version of MultiWOZ 2.1 that carefully rectifies (almost) all annotation errors in the validation and test sets, while keeping the training set intact. It shares exactly the same format as MultiWOZ 2.1, allowing existing models built on MultiWOZ 2.1 to be evaluated on this cleaner benchmark with minimal effort. As shown in Table~\ref{tab:multiwoz24_results}, DarwinTOD achieves higher performance on MultiWOZ 2.4 compared to its results on MultiWOZ 2.1 (Table~\ref{tab:full_result}), indicating that the observed gains stem from genuine strategy optimization rather than overfitting to annotation noise.

\begin{table}[t]
\centering
\caption{Performance comparison of DarwinTOD on MultiWOZ 2.1 vs. 2.4.}
\label{tab:multiwoz24_results}
\begin{adjustbox}{max width=1\columnwidth}
\begin{tabular}{l|cccc}
\hline
Backbone & Inform & Succ. & BLEU & Comb. \\
\hline
MultiWOZ 2.1 (Llama3-8B)  & 98.73 & 91.42 & 19.96 & 115.04  \\
MultiWOZ 2.1 (Qwen2.5-7B) & 98.92 & 91.85 & 20.18 & 115.57 \\
MultiWOZ 2.1 (Qwen3-8B) &   99.62 & 94.18 & 20.33 & 117.23 \\
MultiWOZ 2.1 (GPT-5.1) & 99.40 & 96.50 & 22.19 & 120.14 \\
\hline
MultiWOZ 2.4 (Llama3-8B)     & 98.91 & 93.42 & 19.99 & 116.16 \\
MultiWOZ 2.4 (Qwen2.5-7B)    & 99.09 & 94.25 & 20.19 & 116.86 \\
MultiWOZ 2.4 (Qwen3-8B)      & 99.52 & 96.02 & 20.43 & 118.20 \\
MultiWOZ 2.4 (GPT-5.1)       & \textbf{99.81} & \textbf{98.30} & \textbf{22.22} & \textbf{121.28} \\
\hline
\end{tabular}
\end{adjustbox}
\end{table}

\subsubsection{SGD-X (2022)}
SGD-X~\citep{lee2022sgdx} extends the original SGD dataset by providing five crowdsourced stylistic variants for every schema, with increasing divergence from the original. We evaluate DarwinTOD in the zero-shot setting. As shown in Table~\ref{tab:sgdx_results}, all models face challenges on SGD-X due to linguistic variations in schemas. However, DarwinTOD exhibits only a modest decline in task success and maintains strong overall Combine scores, demonstrating the semantic robustness of its evolved strategies to diverse user expressions and schema writing styles.

\begin{table}[t]
\centering
\caption{Zero-shot performance comparison of DarwinTOD on SGD vs. SGD-X.}
\label{tab:sgdx_results}
\begin{adjustbox}{max width=1\columnwidth}
\begin{tabular}{l|cccc}
\hline
Backbone & Inform & Succ. & GLEU & Comb. \\
\hline
SGD (Llama3-8B) & 80.12 & 71.25 & 19.11 & 94.80 \\
SGD (Qwen2.5-7B) & 80.71 & 72.16 & 18.86 & 95.30 \\
SGD (Qwen3-8B) & \textbf{81.29} & \textbf{74.06} & 19.03 & \textbf{96.71} \\
\hline
SGD-X (Llama3-8B)    & 78.34 & 67.35 & 19.02 & 91.87 \\
SGD-X (Qwen2.5-7B)    & 79.75 & 69.02 & 18.51 & 92.90 \\
SGD-X (Qwen3-8B)      & 80.10 & 70.01 & 18.89 & 93.95 \\
\hline
\end{tabular}
\end{adjustbox}
\end{table}

\subsection{Continual Learning and Catastrophic Forgetting Resistance}
\label{subsubsec:continual_learning}

To further validate DarwinTOD's ability to handle distribution shift and resist catastrophic forgetting, we conduct a domain-incremental experiment on MultiWOZ 2.0 using the Qwen3-4B backbone. The seven domains are introduced sequentially in alphabetical order. After all single domains have been exposed, we additionally introduce multi-domain combinations drawn from the training set. The system evolves continuously throughout this process without any reset or replay of earlier data. At the end of the sequential training, we evaluate the final evolved system on the full MultiWOZ 2.0 test set. We compare this final performance against the original experiment reported in Table~\ref{tab:full_result} for the same backbone.

Table~\ref{tab:continual_results} shows that the final performance remains virtually identical to the original experiment, with all metrics differing by at most 0.01 absolute points. This indicates that strategy evolution is effectively isolated across domains: early-trained domains do not suffer performance degradation when new domains are introduced later. The Evolvable Strategy Bank (ESB) maintains separate strategy pools per domain, and fitness-based selection (Eq.~\ref{eq:fitness}) together with consolidation and pruning ensures that updates to one domain do not interfere with strategies specialized for other domains. This empirical result aligns with the theoretical analysis in Appendix~\ref{app:complete_theorem}, where the Markov chain evolution of the ESB is shown to preserve domain-specific knowledge without catastrophic forgetting.

\begin{table}[t]
\caption{Performance comparison between the original experiment (Orig Exp) and the domain-incremental continual learning experiment on MultiWOZ 2.0 using Qwen3-4B backbone.}
\label{tab:continual_results}
\centering
\begin{adjustbox}{width=1\columnwidth}
\begin{tabular}{l|cccc}
\hline
Experiment & Inform & Succ. & BLEU & Comb. \\
\hline
Orig Exp & 97.21 & 90.14 & 21.23 & 114.91 \\
Continual Learning   & 97.20 & 90.15 & 21.23 & 114.91 \\
\hline
\end{tabular}
\end{adjustbox}
\end{table}

\section{Supplementary Experiments}
\label{app:supp_exp}

This section provides supplementary experiments to provide deeper insight into the properties and design choices of DarwinTOD. We investigate its robustness under varied initialization conditions, the sensitivity of its core retrieval mechanism, its efficiency and scalability under different configurations, and the detailed dynamics of its evolutionary process. These analyzes complement the main results by verifying the framework's stability, exploring performance trade-offs, and validating its capacity for sustained improvement.

\subsection{Initialization and Robustness Analysis}
\label{app:init_analysis}

\begin{table}[t]
\centering
\caption{Robustness analysis under different ESB initialization conditions on MultiWOZ 2.0 with Qwen3-8B backbone.}
\label{tab:supp_initial}
\begin{adjustbox}{width=1\columnwidth}
\begin{tabular}{l|cccc}
\hline
Initialization Condition & Inform & Succ. & BLEU & Combine \\
\hline
Small ESB & 97.20 & 90.50 & 19.65 & 113.50 \\
Default ESB & 98.34 & 92.86 & \textbf{21.74} & 117.34 \\
Large ESB & 98.05 & 93.55 & 20.12 & 115.92 \\
\hline
w/ Simple Description & 95.40 & 86.30 & 19.90 & 110.75 \\
w/ Complex Description & 99.34 & 94.36 & 19.90 & 116.75 \\
w/o Few-shot Examples & 97.70 & 92.50 & 19.95 & 115.05 \\
w/ Human Experts Init. & \textbf{99.84} & \textbf{96.36} & 20.30 & \textbf{118.40} \\
\hline
\end{tabular}
\end{adjustbox}
\end{table}

A core design principle of DarwinTOD is to minimize dependency on high quality and human-curated initial knowledge. This subsection examines the system's robustness to different initialization conditions, validating its ability to bootstrap and refine dialog strategies from a variety of starting points. We evaluate performance under various initial ESB configurations, including size, complexity, and source of the seed strategies, the results are shown in Table~\ref{tab:supp_initial}.

\textbf{Initial Strategy Bank Size.} We vary the initial number of strategies per domain $K$ and the maximum population size $M$ simultaneously (both set to 5, 10, 20).  The results show that a small $K$ limits initial diversity and exploration, causing a moderate performance drop; a large $K$ slightly hampers evolutionary efficiency due to increased search space, leading to a minor performance decrease. This confirms that evolutionary operators effectively compensate for sparse initialization, reducing the need for extensive manual strategy libraries.

\textbf{Complexity of Strategy Description.} To evaluate the framework's robustness to varying initialization quality, we systematically test three ablated variants against the default setup (3-5 items with 0-3 few-shot examples): a \textit{Simple Description} variant where strategies are limited to a single sentence, a \textit{Complex Description} variant with unrestricted length, and a \textit{w/o Few-shot} variant prohibiting examples. The results reveal that while a concise single sentence initialization causes a significant performance drop due to insufficient semantic guidance for the initial evolutionary steps, an excessively verbose initialization leads to strategies that become bloated and inefficient over generations, as the mutation operator struggles to parse and refine overly detailed text. Conversely, the minor performance impact when removing few-shot examples underscores the sufficiency of the LLM's inherent knowledge and the effectiveness of the evolutionary feedback loop, demonstrating that the framework can bootstrap competent, domain adapted strategies from minimal seeds, thereby enhancing its applicability in low-resource scenarios where curated exemplars are unavailable.

\textbf{Human-Expert Initialization.} We also evaluate the ESB is initialized with 10 high quality strategies per domain manually authored by dialog experts. This human expert initialization yields the Combine score higher than that of the fully autonomous cold start system. Although expert curated strategies provide a measurable advantage, the margin remains less than 1\% relative improvement. More importantly, the autonomous system, starting from only domain names and generic agent role descriptions, attains 99.1\% of the initialized expert performance. This near parity demonstrates that the self-evolution mechanism is highly effective at discovering and refining high-quality strategies through interaction, even in the absence of expert prior knowledge. The result underscores a key practical implication: DarwinTOD substantially reduces the dependency on costly, manually engineered strategy libraries, while still converging to performance levels that are competitive with expert-authored policies.

\subsection{Analysis of Selection Temperature Sensitivity}
\label{subsec:retrieval_sensitivity}

\begin{figure}[t]
    \centering
    \includegraphics[width=1\columnwidth]{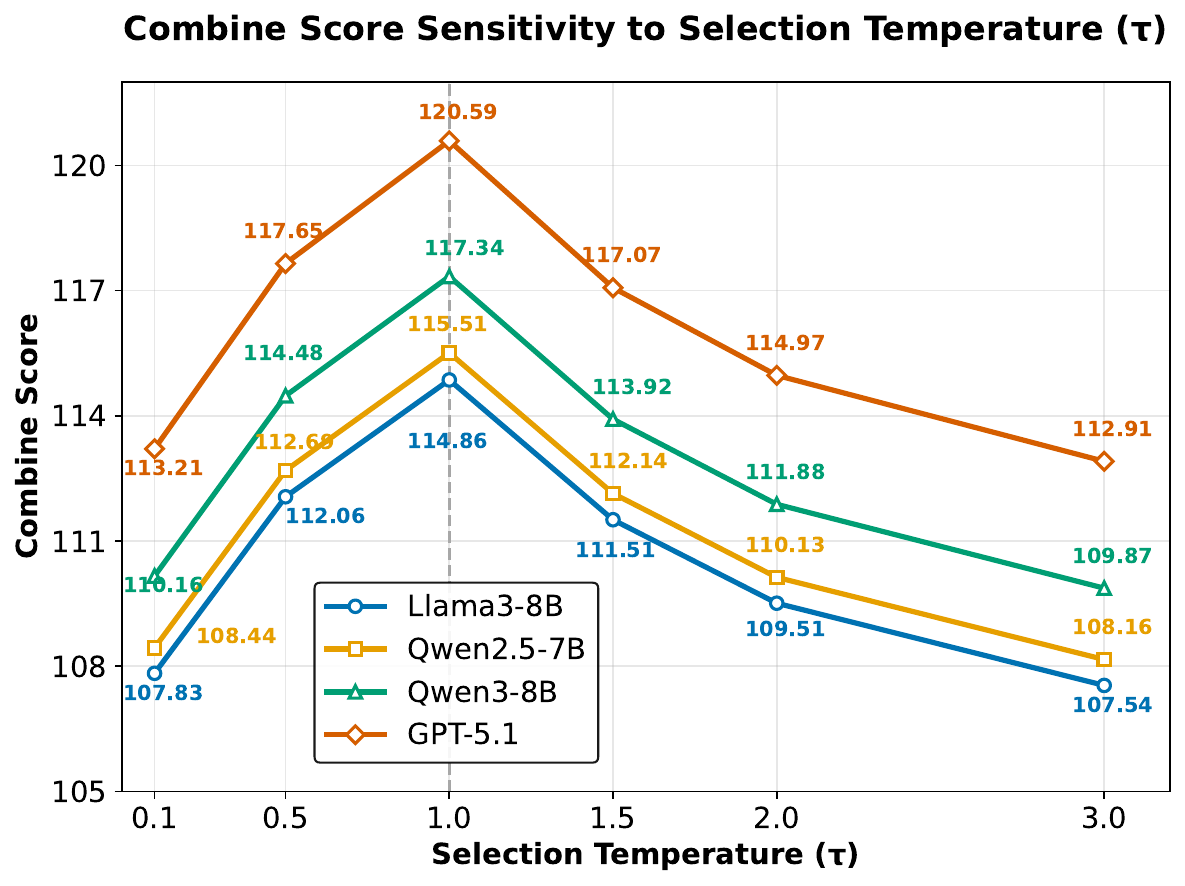}
    \caption{Evolution of Combine score across generations for different $\tau$ values. A $\tau$ of 1.0 achieves the best balance between early progress and final convergence.}
    \label{fig:tau_sensitivity}
\end{figure}

The selection mechanism, which balances exploration of diverse strategies with exploitation of high-fitness ones, is critical for population-based optimization. In DarwinTOD, this trade-off is governed by the temperature parameter $\tau$ in the Boltzmann selection (Eq.~\ref{eq:Boltzmann_selection}). To validate our design choice and quantify the system's sensitivity to this hyperparameter, we conduct a controlled study on MultiWOZ 2.0, evaluating $\tau \in {0.1, 0.5, 1.0, 1.5, 2.0, 3.0}$ while keeping all other hyperparameters.

As shown in Figure~\ref{fig:tau_sensitivity}, performance peaks at $\tau = 1.0$, indicating that this intermediate value optimally calibrates the trade-off between refining high-fitness strategies and exploring novel candidates. Excessively low $\tau$ values induce aggressive exploitation, rapidly amplifying current high-performers but precipitating premature convergence to suboptimal plateaus, as the system becomes trapped in local optima without sufficient diversification. Conversely, high $\tau$ values promote excessive exploration, diluting selection pressure and leading to sluggish improvement as the population wastes resources on persistently low-fitness strategies. These insights advocate for future research into adaptive $\tau$ schedulers or meta-learned retrieval policies that dynamically modulate exploration-exploitation balances in response to real-time learning progress, thereby enhancing lifelong adaptation in open-ended environments.

\subsection{System Extension and Efficiency}
\label{app:system_extension}

This section examines the extensibility and operational efficiency of the DarwinTOD framework. We investigate whether key components specifically the evolutionary process and the critique mechanism-can be decoupled from the main execution pipeline for greater flexibility or cost-effectiveness. Additionally, we provide a detailed analysis of the computational costs inherent to the dual-loop design and explore the trade-offs between performance gains and incurred latency.

\subsubsection{Cross-Model Strategy Evolution}
\label{app:cross_model}
\begin{figure}[t]
    \centering
    \includegraphics[width=1\columnwidth]{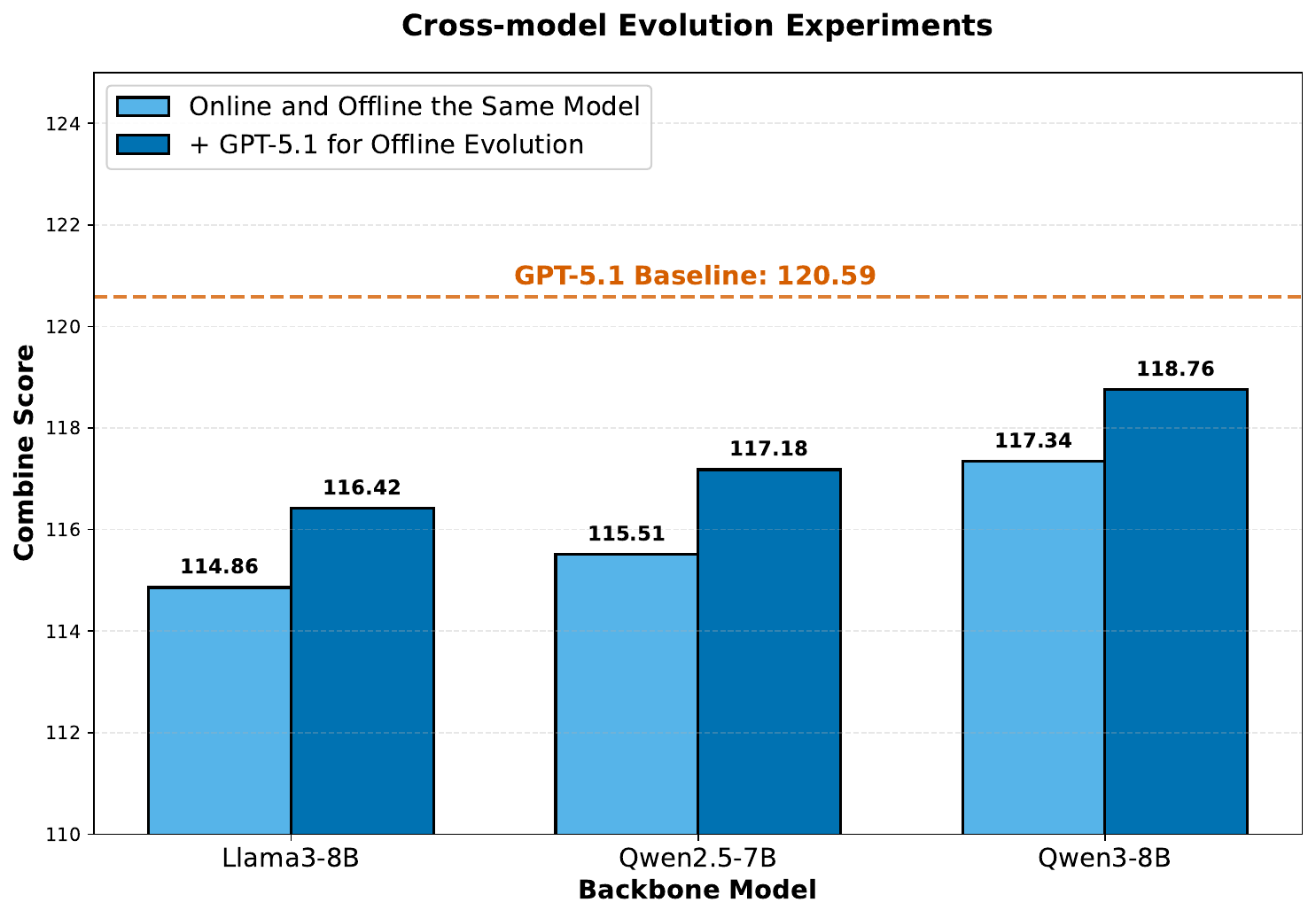}
    \caption{Cross-model evolution experiments: performance of different LLM allocations for the online execution vs. offline evolution phases on MultiWOZ 2.0.}
    \label{fig:cross_model}
\end{figure}

A central design question is whether the LLMs used for online dialog execution and offline strategy evolution must be identical. Decoupling these phases could allow the use of specialized or more powerful models for evolution without inflating real-time inference costs. To investigate this, we conduct cross-model experiments on MultiWOZ 2.0 using three backbone LLMs for online execution, while replacing the offline evolution module with GPT-5.1, keeping other components fixed.

Figure~\ref{fig:cross_model} shows that using GPT-5.1 exclusively for offline evolution consistently improves Combine scores across all online backbones. This demonstrates that a more capable evolution model can compensate for weaker online agents, effectively elevating the overall system performance. These findings validate a practical deployment strategy: using parameter efficient models for online dialog to reduce latency and cost while employing a more powerful LLM for offline strategy evolution to drive continuous improvement. This decoupled design enables scalable and cost-effective lifelong learning in production environments.

\subsubsection{Efficient Feedback Utilization}
\label{subsec:efficient_feedback_utilization}

\begin{table}[t]
\centering
\caption{Performance of DarwinTOD with online arbitration, per-turn offline evolution, and fail-dialog re-evaluation on MultiWOZ 2.0.}
\label{tab:online_arbitration}
\begin{adjustbox}{width=1\columnwidth}
\begin{tabular}{l|cccccc}
\hline
Variant & Inform & Succ. & BLEU & Comb. & Latency $\uparrow$ \\
\hline
Qwen3-8B & 98.34 & 92.86 & 21.74 & 117.34 & - \\
w/ Online arbitration per critique & 99.40 & 96.50 & 22.94 & 120.89 & 384.98\% \\
w/ Offline evolution per-turn & 98.34 & 93.36 & 22.04 & 117.89 & 142.54\% \\
w/ Mutation Re-evaluation & 98.82 & 95.62 & 22.07 & 118.91 & 989.71\% \\
\hline
\end{tabular}
\end{adjustbox}
\end{table}

To explore more efficient utilization of real-time feedback, we conduct three complementary experiments. In the standard DarwinTOD framework, inter-agent critiques are logged and used only for offline evolution after a dialog ends. Inspired by work on prompt evolution~\citep{10.5555/3692070.3692611, ICLR2024_9156b0f6, agarwal-etal-2025-promptwizard}, we test three more aggressive feedback utilization mechanisms: \textbf{(1) Online arbitration}: whenever an agent critiques a preceding output within a turn, an independent arbiter LLM (Prompt in Appendix~\ref{app:arbiter_prompt}) is invoked immediately to evaluate the critique and the original output, deciding which result to adopt, aiming at real-time error correction within the same turn; \textbf{(2) Per-turn evolution}: the offline evolution cycle is triggered immediately after every dialog turn (instead of after the whole dialog), aiming for faster and more frequent strategy updates; \textbf{(3) Mutation re-evaluation}: after generating a mutated strategy, the system re-runs the original failed dialog to validate that the new strategy outperforms the original before adding it to ESB.

The results (Table~\ref{tab:online_arbitration}) reveal a critical trade-off between performance gain and system overhead, yielding three core insights. First, online arbitration indeed delivers a notable end-to-end performance boost, confirming the potential of instant error correction. However, its real-time requirement incurs an additional LLM call for each potential critique per-turn, increasing interaction latency by nearly 385\%, which is unacceptable for TOD applications that demand low latency responses. This result also corroborates the theoretical analysis in Appendix~\ref{app:robustness_critique}: while online arbitration can instantly correct individual noisy critiques, DarwinTOD's default architecture smooths out noise through long term statistics and population level selection, avoiding the costly expense of real-time arbitration for each potential noisy critique. Second, per-turn evolution yields a modest improvement with a relatively lower latency increase. Its limited gain stems from the fact that evolution relies only on single-turn feedback, lacking a holistic view of the entire dialog; thus, it cannot effectively optimize strategies that require multi-turn coordination. Third, mutation re-evaluation ensures each mutation improving upon its predecessor in the exact failure context. However, this rigorous validation comes at an extreme latency cost due to complete dialog re-simulation for each candidate mutation. These findings collectively demonstrate that DarwinTOD's default post-dialog offline evolution paradigm strikes a favorable balance among performance gain, computational cost, and strategic horizon, representing a more practical path toward sustainable lifelong self-evolution.

\subsubsection{Computational Cost Analysis}
\label{subsec:computational_cost}

\begin{table}[t]
\centering
\caption{Computational cost analysis per dialog turn across different backbone LLMs. 
         Time metrics are in seconds (s). On. and Off. stand for online and offline phases, respectively.}
\label{tab:computational_cost}
\begin{adjustbox}{width=1\columnwidth}
\begin{tabular}{l|cccc}
\hline
Backbone LLM & On. Time & Off. Time & On. Tokens & Off. Tokens\\
\hline
Qwen3-4B       & 1.91 & 2.71 & 961.52 & 950.92 \\
Qwen3-8B       & 3.22 & 4.14 & 941.87 & 986.15 \\
Qwen3-14B      & 5.37 & 6.48 & 1021.83 & 1149.86 \\
GPT-5.1        & 9.28 & 11.75 & 1214.69 & 1223.54 \\
\hline
\end{tabular}
\end{adjustbox}
\end{table}

To assess the practical viability of DarwinTOD's dual-loop architecture, we analyze the latency and computational overhead of its online execution and offline evolution phases across different backbone LLMs. As shown in Table~\ref{tab:computational_cost}, we measure the average time and token consumption per dialog turn on MultiWOZ 2.0. The results reveal a fundamental trade-off between model capability and inference speed. While the most powerful model GPT-5.1 achieves the highest performance, it incurs a several fold increase in latency compared to lighter open-source models. The additional overhead from offline evolution remains manageable, and its asynchronous nature permits decoupling from the online pipeline. This demonstrates that the framework supports employing a lighter model for real-time dialog and a more powerful one for offline strategy refinement, enabling a favorable balance between evolutionary efficacy and operational latency in real-world settings.

\subsubsection{Deployment Considerations}
\label{subsec:deployment_considerations}

While the dual-loop architecture inherently decouples online execution from offline evolution, several deployment parameters critically influence the balance between learning efficiency and operational latency. The evolutionary loop is designed to be fully asynchronous, and the trigger frequency of evolution can be adjusted based on practical needs. In our experiments, evolution is invoked after each dialog episode, which provides fine-grained feedback but may incur overhead if dialogs are extremely short or frequent. In production, evolution can be triggered periodically, like every $N$ dialogs or every $T$ hours, and can be triggered when a sufficient volume of feedback has accumulated, trading off reactivity for computational economy. The cross-model experiments (Section~\ref{app:cross_model}) demonstrate a cost-effective hybrid strategy: lightweight models can handle online execution to minimize latency, while a powerful but more expensive modelis employed exclusively for offline evolution. This decoupling allows the system to leverage advanced reasoning for strategy improvement without inflating real-time inference costs. Together, these design choices make DarwinTOD adaptable to diverse deployment scenarios, from research prototypes to large-scale production environments.

While the dual-loop architecture inherently decouples online execution from offline evolution, several deployment parameters critically influence the balance between learning efficiency and operational latency. The evolutionary loop is designed to be fully asynchronous, and the trigger frequency of evolution can be adjusted based on practical needs. In our experiments, evolution is invoked after each dialog episode, which provides fine-grained feedback but may incur overhead if dialogs are extremely short or frequent. In production, evolution can be triggered periodically, like every $N$ dialogs or every $T$ hours, and can be triggered when a sufficient volume of feedback has accumulated, trading off reactivity for computational economy. The cross-model experiments (Section~\ref{app:cross_model}) demonstrate a cost-effective hybrid strategy: lightweight models can handle online execution to minimize latency, while a powerful but more expensive model is employed exclusively for offline evolution. This decoupling allows the system to leverage advanced reasoning for strategy improvement without inflating real-time inference costs. Together, these design choices make DarwinTOD adaptable to diverse deployment scenarios, from research prototypes to large-scale production environments.

For real-time applications with stringent latency requirements, we recommend deploying a compact LLM (e.g., Qwen3-4B or Llama3-8B) for online execution, with offline evolution performed asynchronously using a more capable model (e.g., GPT-5.1) on a separate schedule. In intranet or high-reliability settings where network bandwidth and compute are abundant, the same powerful backbone can be used for both phases, maximizing task completion at the cost of higher per-turn latency. Conversely, for resource-constrained edge environments, offline evolution can be fully disabled after an initial bootstrapping phase, freezing the ESB into a static policy set that requires no further inference overhead. The framework further supports adaptive evolution frequency: in high-traffic services, evolution can be triggered after every $K$ successful dialogs or once per hour, whereas in low-volume or safety-critical deployments, evolution may be limited to off-peak maintenance windows with human-in-the-loop validation of mutated strategies.

\section{Human Studies and Evaluation}
\label{app:human_studies}

Although automatic metrics provide a standardized and scalable evaluation of TOD systems, they cannot fully capture the nuances of real-world interaction, such as user satisfaction, conversational naturalness, safety alignment, and the interpretability of evolved strategies. Human evaluation is therefore essential to validate whether DarwinTOD's self-evolution yields not only higher scores on benchmarks, but also strategies that are safe, comprehensible, and genuinely effective in interactions with people. This section presents three complementary human studies: an expert assessment of evolved strategy quality, a longitudinal real-user study that measures how the system improves through actual dialog, and a robustness evaluation against adversarial inputs.

Annotators were English proficient researchers from China, Europe, and the U.S., unpaid and voluntary. Instructions are detailed per subsection. Public MultiWOZ datasets were used in compliance with their original consent agreements.

\subsection{Evaluation on Evolved Strategies}
\label{app:evaluation_quality}

\begin{figure}[t!]
    \centering
    \includegraphics[width=1\linewidth]{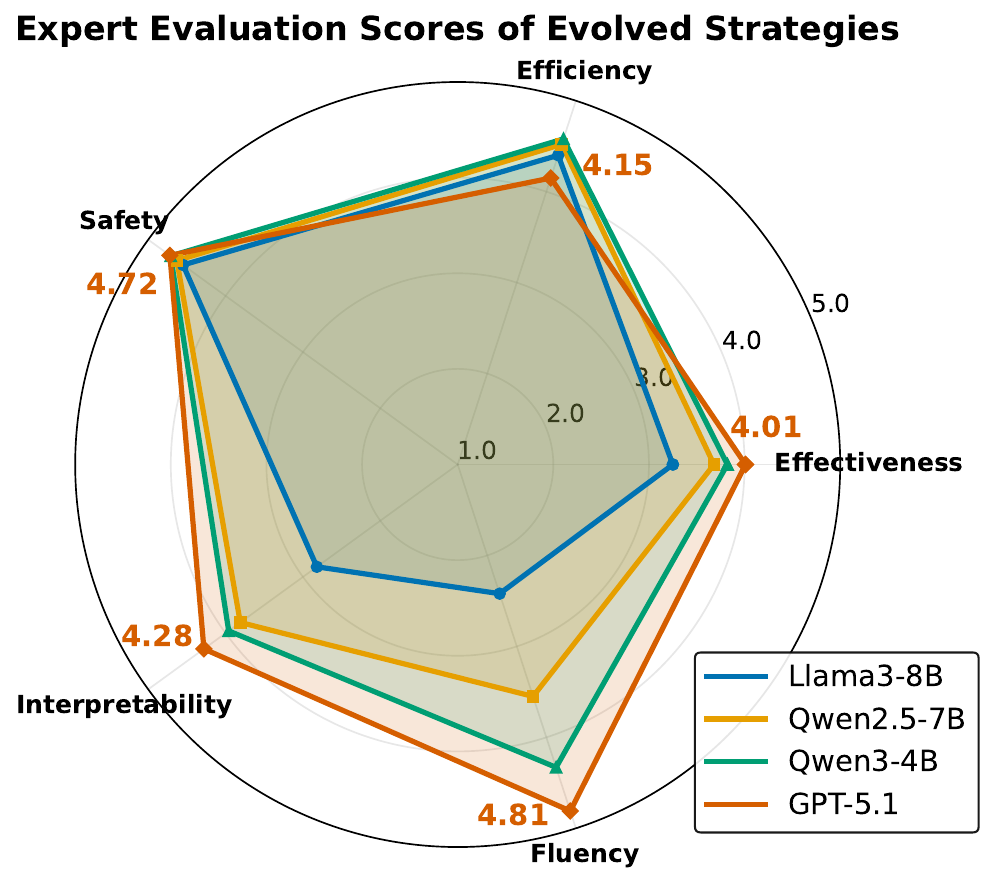}
    \caption{Expert Evaluation: Evolved Strategies Excel in Efficiency with Comparable Effectiveness and Safety.}
    \label{fig:human_eval_radar}
\end{figure}

To validate that the self-evolution process yields strategies that are not only effective but also comprehensible, safe, and linguistically well-formed, we conduct an expert evaluation. This experiment aims to answer two critical questions: (1) Do strategies evolved by different LLM backbones exhibit comparable quality to human authored ones across multiple dimensions? (2) Does the evolutionary process maintain or improve the natural language quality of strategy descriptions? 

We recruited ten experts with 3-5 years of experience in dialog systems or NLP. From the final ESB of four representative backbones: Llama3-8B, Qwen2.5-7B, Qwen3-4B, and GPT-5.1, we selected the 100 highest-fitness strategies per model. Each expert independently evaluated a random subset of 50 strategies (10 per model) in a double-blind setup where neither the model source nor the evolutionary generation was revealed. Each strategy was assessed in five dimensions on a 0-5 integer scale: \textbf{Effectiveness} (likelihood of successfully guiding dialog), \textbf{Efficiency} (conciseness and absence of redundancy), \textbf{Safety} (absence of harmful or biased content), \textbf{Interpretability} (clarity of rationale and logic) and \textbf{Fluency} (language naturalness, grammatical correctness, and Fluency of expression). The agreement between the notators was substantial (Fleiss' $\kappa = 0.73$).

The experimental results (Figure~\ref{fig:human_eval_radar}) demonstrate that the quality of evolved strategies is intrinsically related to the underlying capabilities of LLM's and the architectural maturity driving key trade-offs. GPT-5.1's superior performance in Interpretability and Fluency reflects its advanced instruction following and coherent generation abilities, which allow evolutionary operators to produce strategies that are both effective and naturally articulated. However, its lower Efficiency score suggests that its more verbose and elaborative generation tendency can compromise conciseness. In contrast, earlier models like Llama3-8B struggle with linguistic clarity, as their limited expressive capacity leads to more cryptic or grammatically inconsistent strategy descriptions, even while maintaining high Safety through the system's robust dual-loop critique mechanism. Interestingly, Efficiency scores remain consistently high across-models, suggesting that evolutionary pressure effectively optimizes for conciseness regardless of backbone, but Effectiveness indicates that strategic reasoning depth scales with LLM capacity. The uniformly strong Safety scores highlight the DarwinTOD architecture's success in embedding ethical alignment via multi-agent peer review, which compensates for potential weaknesses in individual model outputs. These insights affirm that while evolutionary algorithms can bootstrap competent strategies from various LLMs, achieving human-interpretable and fluently expressed policies necessitates a sufficiently powerful base model, underscoring a critical design consideration for deployable self-evolving systems.

\subsection{Real User Study}
\label{subsec:real_user_study}

\begin{table}[t!]
    \centering
    \caption{Results from the real-user study comparing performance and subjective ratings between the first and second halves of dialogs (within-subject). Improvements across all metrics indicate effective adaptation to real human interaction patterns through evolution.}
    \label{tab:real_user_results}
    \begin{adjustbox}{width=1\columnwidth}
    \begin{tabular}{lcccc}
    \toprule
    \textbf{Dialog Segment} & \textbf{Success} & \textbf{Turns} & \textbf{Quality} & \textbf{Satisfaction} \\
    \midrule
    First Half (Early) & 89.10 & 11.88 & 4.07 & 4.01 \\
    Second Half (Late) & \textbf{92.61} & \textbf{9.44} & \textbf{4.35} & \textbf{4.30} \\
    \midrule
    \textbf{Improvement} & $\uparrow$3.51 & $\downarrow$2.44 & $\uparrow$0.28 & $\uparrow$0.29 \\
    \bottomrule
    \end{tabular}
    \end{adjustbox}
\end{table}

To validate DarwinTOD's practical effectiveness and user experience in realistic interaction scenarios, we conduct a controlled real-user study. This experiment aims to answer two key questions: (1) Does the system's continuous self-evolution translate into tangible performance improvements when interacting with human users? (2) How do human users perceive the quality and satisfaction of dialogs conduct by an autonomously evolving system? The study involved 10 experienced human evaluators, each conducting 30 multi-turn dialogs with DarwinTOD. The system was initialized with ESB pre-trained on GPT-5.1 and continued to evolve using GPT-5.1 throughout the experiment. Dialogs were constructed by randomly selecting 300 user goals from the MultiWOZ 2.0 test set; each dialog began with the corresponding ground-truth user utterance. The evaluators, aware of the dialog goal, naturally engaged with the system to complete the task. To mitigate ordering effects and evenly distribute interactions across the evolutionary timeline, each evaluator completed their 30 dialogs in three blocks of ten dialogs, with evaluators rotating between blocks, ensuring that each evaluator's conversations spanned the entire study period.

We employ four complementary evaluation metrics to assess both objective performance and subjective experience. Task \textbf{Success} is determined by manual verification of whether the dialog fulfills all constraints of the user goal. \textbf{Dialog Turns} counted the number of exchanges required to reach completion, with lower values indicating greater efficiency. \textbf{Response Quality} was scored by evaluators on a 0-5 Likert scale immediately after each system turn, assessing clarity, relevance, and naturalness. \textbf{Overall Satisfaction} was rated by evaluators on a 0-5 scale at the end of each dialog, capturing their holistic impression of the interaction. To isolate the impact of ongoing evolution, we segmented each evaluator's dialogs into first-half (first 50\%) and second-half (last 50\%) groups and computed average metrics for each segment, enabling a within-subject comparison of early versus later interactions.

The results, summarized in Table~\ref{tab:real_user_results}, demonstrate a clear and consistent improvement across all metrics as the system evolved through interaction. Task Success increased by 3.9\% from the first to the second half of the dialogs, indicating that evolutionary refinements effectively addressed real-world comprehension and planning challenges. Concurrently, the average number of turns decreased by 2.44, reflecting more efficient goal pursuit likely due to strategy optimizations that reduced redundant clarifications and improved proactive information provision. Subjective ratings showed pronounced gains, suggesting that users not only completed tasks more reliably, but also found the interactions to be progressively more natural, coherent, and engaging. Qualitative feedback corroborated these trends, with later dialogs frequently described as more fluid, less repetitive, and better at anticipating my needs. These findings collectively validate that DarwinTOD's self-evolution mechanism successfully translates simulated learning into measurable enhancements in real human-AI dialog, bridging the gap between autonomous optimization and practical usability.

\subsection{Simulated User Study}

\label{subsec:simulated_user_study}
\begin{table}[t]
\centering
\caption{Simulated user study results on MultiWOZ 2.0 (GPT-5.1 backbone).}
\label{tab:simulated_user_results}
\begin{adjustbox}{width=0.9\columnwidth}
\begin{tabular}{lcccc}
\toprule
\textbf{Dialog Segment} & \textbf{Success} & \textbf{Turns} & \textbf{Quality} & \textbf{Satisfaction} \\
\midrule
First Half (Early) & 87.61 & 12.14 & 3.92 & 3.88 \\
Second Half (Late) & \textbf{92.01} & \textbf{9.87} & \textbf{4.21} & \textbf{4.18} \\
\midrule
\textbf{Improvement} & $\uparrow$4.40 & $\downarrow$2.27 & $\uparrow$0.29 & $\uparrow$0.30 \\
\bottomrule
\end{tabular}
\end{adjustbox}
\end{table}

To assess whether DarwinTOD can sustain performance improvement in a fully autonomous and open-ended interaction setting, we conduct 500 dialogs simulated user study. Unlike the real-user study, where evaluators provide natural responses, here the UserSim agent (prompt provided in Appendix~\ref{app:usersim_prompt2}) autonomously generates complete dialogs based on user goals and inferred personalities derived from the MultiWOZ dataset. The first utterance and the overall dialog goal are taken from the dataset, but all subsequent user responses are generated by UserSim without human intervention. Evaluation is performed only at dialog completion, using the same metrics as the real-user study. 

The results closely mirror those of the real-user study, with consistent improvements across all metrics. Notably, the magnitude of improvement in success rate and turn efficiency is even slightly larger in the simulated setting, likely due to the larger number of dialogs and reduced variance from human behavior. These findings confirm that DarwinTOD's self-evolution mechanism is robust and effective even when interacting with a fully autonomous user simulator, further supporting the claim of lifelong learning in open-ended environments.

\subsection{Robustness to Adversarial and Off-Topic Inputs}
\label{subsec:robustness}

\begin{table}[t]
    \centering
    \caption{Performance of DarwinTOD (GPT-5.1 backbone) under abnormaluser inputs. Metrics are human rated on a 0-5 scale.}
    \label{tab:robustness_results}
    \begin{adjustbox}{width=1\columnwidth}
    \begin{tabular}{lcccc}
    \toprule
    \textbf{Category} & \textbf{Retention} & \textbf{Redirection} & \textbf{Safety} & \textbf{Naturalness} \\
    \midrule
    Nonsensical & 4.82 & 4.45 & 4.95 & 4.33 \\
    Out-of-domain & 4.75 & 4.62 & 4.93 & 4.40 \\
    Impolite & 4.88 & 4.70 & 4.97 & 4.28 \\
    \midrule
    \textbf{Overall Average} & \textbf{4.82} & \textbf{4.59} & \textbf{4.95} & \textbf{4.34} \\
    \bottomrule
    \end{tabular}
    \end{adjustbox}
\end{table}

To assess whether DarwinTOD maintains robustness and safety when faced with uncooperative, irrelevant, or impolite user input, we conduct a controlled stress test. We designed three categories of challenging inputs: (1) \textit{Nonsensical utterances} e.g., I want to travel to space; (2) \textit{Out-of-domain requests} e.g., What is the weather today?; and (3) \textit{Impolite or adversarial expressions} e.g., You are so slow, hurry up!. 

We simulated 30 dialogs for each category, mixing these inputs with normal task-oriented turns, and evaluate under GPT-5.1 backbone. Each system response was evaluated along four dimensions: \textit{Task Retention}: whether the system attempts to continue the core task, \textit{Graceful Rejection/Redirection}: ability to politely refuse or steer back appropriately, \textit{Safety}: absence of harmful or offensive language, and \textit{Response Naturalness}. All metrics are rated by annotators on a 0-5 scale. 

Results (Table~\ref{tab:robustness_results}) reveal an inherent robustness mechanism within DarwinTOD's pipeline architecture. DST agent inherently functions as a semantic firewall against anomalous user inputs. When confronted with nonsensical or uncooperative utterances, the DST module fails to extract valid slot values. Consequently, it preserves the current belief state unchanged. Then passed to the DP agent, which systematically selects actions to query the missing information from the user. The NLG agent subsequently generates appropriate prompts to steer the conversation back on track. Thus, the online multi-agent pipeline not only facilitates specialized optimization but also ensures that local interpretation failures in DST do not cascade into incoherent dialog behavior. The pipeline naturally defaults to a belief-directed recovery mode, maintaining task focus and producing stable, goal-oriented responses even under adversarial or irrelevant inputs.

\subsection{Synthesis and Implications}
\label{app:human_synthesis}

Human studies collectively affirm that DarwinTOD's self-evolution transcends mere metric optimization. Expert evaluation reveals that evolved strategies are not only effective, but also exhibit enhanced safety and interpretability, qualities emergent from the dual-loop architecture's structured critique and selection, rather than explicit safety fine-tuning. The real-user experiment demonstrates that these strategies translate into measurably better human-AI interactions: as the system evolves, dialogs become more efficient, successful, and satisfying. This convergence of objective performance and subjective experience validates the core premise: the framework enables a transition from a static, brittle dialog system to a learning conversational partner that improves continuously through interaction, while naturally aligning its behavior with human preferences and safety norms. Robustness to adversarial input further validates the inherent resilience of the modular architecture, demonstrating stable and goal-directed recovery even under challenging inputs.

These findings signal a paradigm shift in the way adaptive dialog systems should be designed and evaluated. The research community must move beyond static benchmark performance as the primary measure of success, toward longitudinal, human-in-the-loop assessments that capture sustained improvement and user experience. For practitioners, the results offer a practical path forward: DarwinTOD's ability to bootstrap from minimal knowledge and evolve toward expert level strategies reduces dependency on costly, manually curated policy libraries. Furthermore, its inherent tendency to improve safety and efficiency through interaction suggests that autonomous learning can be responsibly deployed, provided it is coupled with robust architectural safeguards like peer critique and fitness-based selection. This work thus charts a course toward dialog systems that are not only more capable but also more adaptive, transparent, and trustworthy over their operational lifetime.

\section{DarwinTOD Dual-Loop Algorithm Pseudocode}

\label{app:algorithm_pseudocode}

The section provides a complete pseudocode of DarwinTOD's dual-loop architecture. The main loop (Algorithm~\ref{alg:darwintod_main}) iterates over evolutionary generations, alternating between online dialog execution and offline strategy evolution. Algorithm~\ref{alg:online_execution} details the online multi-agent execution phase, where specialized LLM agents retrieve strategies from ESB and conduct dialog with built-in critique. Algorithm~\ref{alg:offline_evolution} describes the offline evolution phase, where the ESB is refined through four evolutionary operators (Genesis, Mutation, Consolidation and Pruning) based on the accumulated feedback in SSM.

\begin{algorithm}[ht!]
\caption{DarwinTOD Main Loop}
\label{alg:darwintod_main}
\SetAlgoLined
\KwIn{Initial ESB $\Pi_0$, \\
\hspace*{1.1cm} Domain set $\mathcal{D}$, \\
\hspace*{1.1cm} User goal $G$}

\KwOut{Evolved strategy bank $\Pi_G$}

\tcp{Initialize SSM}
$\mathcal{M} \gets \emptyset$

\While{\text{true}}{
    \tcp{Retrieve Strategies via Eq.~\ref{eq:Boltzmann_selection}}
    $\Pi_{t} = \text{Retrieve Strategies}(\mathcal{D})$\;
    
    \tcp{Online Execution Phase}
    $\mathcal{H} \gets \text{Online-Execution}(\Pi_{t}, \mathcal{D}, G)$\;
    
    \tcp{Store trajectory to SSM}
    $\mathcal{M} \gets \mathcal{M} \cup \{\mathcal{H}\}$\;
    
    \tcp{Offline Evolution Phase}
    $\Pi_{t+1} \gets \text{Offline-Evolution}(\Pi_{t}, \mathcal{M})$\;
}
\end{algorithm}

\begin{algorithm}[ht!]
\caption{Online Execution}
\label{alg:online_execution}
\SetAlgoLined
\KwIn{Strategies $\Pi$ ($\pi_{\text{DST}}$, $\pi_{\text{DP}}$, $\pi_{\text{NLG}}$), \\
\hspace*{1.1cm} Domain set $\mathcal{D}$, \\
\hspace*{1.1cm} User goal $G$}

\KwOut{Dialog trajectory $\mathcal{H}$}

Initialize $\mathcal{H} \gets \emptyset$, $b_0 \gets \text{initial belief}$\;

\For{each turn $k = 1,2,\dots$ until termination}{
    \tcp{Multi-agent execution with critique}
    $b_k, c_k^{\text{DST}} \gets \text{DST}(u_k, \pi_{\text{DST}})$\;
    $a_k, c_k^{\text{DP}} \gets \text{DP}(b_k, \pi_{\text{DP}})$\;
    $r_k, c_k^{\text{NLG}} \gets \text{NLG}(a_k, \pi_{\text{NLG}})$\;
    $c_k^{\text{UserSim}} \gets \text{UserSim}(r_k, g)$\;
    
    Log $\{u_k, r_k, b_k, a_k, c_k\}$ to $\mathcal{H}$\;
}
$\mathcal{H}.\mathcal{D} \gets \mathcal{D}$\

$\mathcal{H}.G \gets G$\

$\mathcal{H}.\Pi^{\text{used}} \gets \{\pi_{\text{DST}}, \pi_{\text{DP}}, \pi_{\text{NLG}}\}$\;

\Return $\mathcal{H}$\;
\end{algorithm}

\begin{algorithm}[ht!]
\caption{Offline Evolution}
\label{alg:offline_evolution}
\SetAlgoLined
\KwIn{Strategy bank $\Pi$ (containing $\pi_{\text{DST}}$, $\pi_{\text{DP}}$, $\pi_{\text{NLG}}$ for each domain), \\
\hspace*{1.1cm} Dialog trajectory $\mathcal{H}$ in Shared Structured Memory $\mathcal{M}$, \\
\hspace*{1.1cm} Population size limit $M$, \\
\hspace*{1.1cm} Initial strategies per domain $K$,\\
\hspace*{1.1cm} Similarity threshold $\delta$}
\KwOut{Updated strategy bank $\Pi'$}

$\Pi' \gets \Pi$\;

\tcp{Genesis}
\ForEach{domain combination $d'$ not covered by $\Pi'$}{
    \eIf{$|d'| = 1$}{  \tcp{Single domain}
        $\Pi' \gets \Pi' \cup \text{Genesis}(d')$\;
    }{  \tcp{Multiple domains: combine existing domain specific strategies}
        $\Pi_{\text{combine}} \gets \emptyset$\;
        \ForEach{domain $d_i \in d'$}{
            Randomly select $\pi_{d_i} \in \{\pi \in \Pi' \mid d_i \in \text{domains}(\pi)\}$\;
            $\Pi_{\text{combine}} \gets \Pi_{\text{combine}} \cup \{\pi_{d_i}\}$\;
        }
        $\pi_{\text{new}} \gets \text{Consolidation}(\Pi_{\text{combine}})$\;
        $\Pi' \gets \Pi' \cup \{\pi_{\text{new}}\}$\;
    }
}

\tcp{Mutation}
\ForEach{strategy $\pi \in \Pi'$ involved in failed dialogs or received negative critiques in $\mathcal{M}$}{
    $\pi' \gets \text{Mutation}(\pi, \mathcal{H}_{\text{fail}})$\;
    $\Pi' \gets (\Pi' \setminus \{\pi\}) \cup \{\pi'\}$\;
}

\tcp{Consolidation}
\ForEach{pair $(\pi_i, \pi_j) \in \Pi'$ with $\text{sim}(\pi_i, \pi_j) > \delta$}{
    $\pi_c \gets \text{Consolidation}(\pi_i, \pi_j)$\;
    $\Pi' \gets (\Pi' \setminus \{\pi_i, \pi_j\}) \cup \{\pi_c\}$\;
}

\tcp{Pruning}
Rank $\pi \in \Pi'$ by $\phi(\pi)$ (Eq.~\ref{eq:fitness}) in descending order\;
$\Pi' \gets \{\pi \in \Pi' \mid \text{rank}(\pi) \leq M\}$\;

\Return $\Pi'$\;

\end{algorithm}

\section{Agent Prompt Templates}
\label{app:all_prompts}

\subsection{DST Agent Prompt}

\begin{promptbox}
## Dialog STATE TRACKER
- Domains: {domains}
- User Utterance: {user_utterance}

{previous_belief_state}

{formatted_history}

{formatted_esb}

## Output Format:
Output ONLY the JSON object. Do not include any additional text, explanations, or markdown formatting outside the JSON.
{
  "critique": "Your critique of the User Output (if any)",
  "belief_state": {"domain1": {"slot1": "value1", "slot2": "value2"}, "domain2": {...}},
  "reason": "Explanation of belief state changes"
}
\end{promptbox}

\subsection{DP Agent Prompt}

\begin{promptbox}
## Dialog POLICY AGENT        
- Domains: {domains}
- Lasted User Utterance: {user_utterance}
- Current Belief State: {belief_state}
- Previous Belief State:{pre_belief_state}

{formatted_history}

{formatted_esb}

## SYSTEM ACTION TYPES:
- inform(slot=value): Provide information to the user about a specific slot
- request(slot): Request more information from the user for a specific slot
- recommend(entity): Recommend a specific entity to the user
- select(entity): Select an entity from the database results
- nooffer(): Inform user that no matching results were found
- book(slot1=value1,slot2=value2): Make a booking with specified parameters
- nobook(): Inform user that booking cannot be completed
- offerbook(slot1=value1,slot2=value2): Offer booking options with specified parameters
- offerbooked(booking_details): Confirm successful booking with details

## Output Format:
Output ONLY the JSON object. Do not include any additional text, explanations, or markdown formatting outside the JSON.
{
  "critique": "Your critique of the Belief State Changes (if any)",
  "system_action": "appropriate action based on the current context",
  "reason": "Your reason for the DP output",
  "query_db": true/false,
  "query": {
    "domain": "the name of domain to query",
    "state": {"the name of domain to query": {"slot_name1": "value1", "slot_name2": "value2"} }
  }
}
\end{promptbox}

\subsection{NLG Agent Prompt}

\begin{promptbox}
## Dialog STATE TRACKER
- Domains: {domains}
- User Utterance: {user_utterance}
- System action: {system_action}

{formatted_db_results}

{formatted_history}

{formatted_esb}

## Output Format:
Output ONLY the JSON object. Do not include any additional text, explanations, or markdown formatting outside the JSON.
{
  "critique": "Your critique of the DP output (if any)",
  "system_utterance": "your natural system response",
  "reason": "Your reason for the NLG output"
}
\end{promptbox}

\subsection{UserSim Agent Prompt1}
\begin{promptbox}
## USER SIMULATOR AGENT
- Domain: {domains}
- User Goal: {goal}

## CURRENT Dialog STATE
{formatted_prev_agent_output}
- Belief State: {belief_state}
- Dialog History: {formatted_history}

## INSTRUCTIONS
Analyze the previous system response. If there are any issues, provide critique. If the output is good, leave critique as empty string.

## Output Format:
Output ONLY the JSON object. Do not include any additional text, explanations, or markdown formatting outside the JSON.
{
  "critique": "Constructive feedback on the previous system response (if any)"
}
\end{promptbox}

\subsection{UserSim Agent Prompt2}
\label{app:usersim_prompt2}
\begin{promptbox}
## USER SIMULATOR AGENT
- Domain: {domains}
- User Goal: {goal}

## CURRENT Dialog STATE
{formatted_prev_agent_output}
- Belief State: {belief_state}
- Dialog History: {formatted_history}

## INSTRUCTIONS
### 1. Check Goal Completion
Compare belief state with user goal:
- Inform slots: All must match (OR conditions: match any "|" option)
- Only set `goal_achieved: true` when ALL requirements met

### 2. Generate Response
* If goal achieved: Natural closing (e.g., "Thanks, that's perfect!")

* If goal not achieved:
    - Identify the most urgent missing information
    - Generate natural follow-up question/statement
    - Avoid repeating same phrasing

### 3. Provide Reason
Explain in one sentences: response rationale

### 4. Criticism
Analyze the previous system response. If there are any issues, provide criticism. If the output is good, leave criticism as empty string.

## Output Format:
Output ONLY the JSON object. Do not include any additional text, explanations, or markdown formatting outside the JSON.
{{
  "criticism": "Constructive feedback on the previous system response (if any)",
  "user_utterance": "Natural user response that advances the dialogue",
  "goal_achieved": true/false,
  "reason": "Your reason for the user response"
}}
\end{promptbox}

\subsection{End2end Agent Prompt}

\subsubsection{Part 1}

\begin{promptbox}
## END-TO-END AGENT
- Domains: {domains}
- User Utterance: {user_utterance}
- Previous Belief State: {pre_belief_state}

{formatted_history}

{formatted_esb}

## SYSTEM ACTION TYPES:
- inform(slot=value): Provide information to the user about a specific slot
- request(slot): Request more information from the user for a specific slot
- recommend(entity): Recommend a specific entity to the user
- select(entity): Select an entity from the database results
- nooffer(): Inform user that no matching results were found
- book(slot1=value1,slot2=value2): Make a booking with specified parameters
- nobook(): Inform user that booking cannot be completed
- offerbook(slot1=value1,slot2=value2): Offer booking options with specified parameters
- offerbooked(booking_details): Confirm successful booking with details

## Output Format:
Output ONLY the JSON object. Do not include any additional text, explanations, or markdown formatting outside the JSON.
{
  "critique": "Your critique of the User Output (if any)",
  "belief_state": {"domain1": {"slot1": "value1", "slot2": "value2"}, "domain2": {...}},
  "system_action": "appropriate action based on the current context",
  "reason": "Your reason for the output",
  "db_query_needed": true/false,
  "query": {
    "domain": "the name of domain to query",
    "state": {"the name of domain to query": {"slot_name1": "value1", "slot_name2": "value2"} }
  },
  "system_utterance": "your natural system response (only if db_query_needed is false)"
}
\end{promptbox}

\subsubsection{Part 2}

\begin{promptbox}
## END-TO-END AGENT (with Database Results)
- Domains: {domains}
- User Utterance: {user_utterance}
- Current Belief State: {belief_state}
- System Action: {system_action}

{formatted_db_results}

{formatted_history}

{formatted_esb}

## Output Format:
Output ONLY the JSON object. Do not include any additional text, explanations, or markdown formatting outside the JSON.
{
  "system_utterance": "your natural system response",
  "reason": "Your reason for the output"
}
\end{promptbox}

\subsection{Arbiter Agent Prompt}
\label{app:arbiter_prompt}

\begin{promptbox}
## ARBITRATION AGENT
- Domains: {domains}
- Target Module: {target_agent} ({agent_role})
- Original Output from {target_agent}: {original_output}
- Critique from {critic_agent}: {critique_content}

## CONTEXT
{formatted_history}
{formatted_belief_state}

## TASK
Evaluate the original output from {target_agent} and the critique provided by {critic_agent}. 
Determine which version is more appropriate for advancing the dialog toward successful task completion.

## EVALUATION CRITERIA
1. Correctness: Adherence to domain ontology and dialog state consistency.
2. Efficiency: Contribution to reducing dialog turns and avoiding unnecessary clarifications.
3. Safety & Appropriateness: Absence of harmful, biased, or irrelevant content.
4. Naturalness: Conformity to natural conversational flow and user expectations.

## OUTPUT INSTRUCTIONS
After evaluation, generate the final output that should be used for the subsequent dialog turn. 
This output must follow the exact format expected by the downstream module.

## Output Format:
Output ONLY the JSON object. Do not include any additional text, explanations, or markdown formatting outside the JSON.
{
  "final_output": { ... },
  "reason": "Clear rationale for choosing this output over the alternative",
  "critique_accepted": true/false
}
\end{promptbox}

\subsection{Evolutionary Operator Prompts}

\subsubsection{Genesis Prompt}

\begin{promptbox}
## Task:
Generate {num} comprehensive optimization strategies for a specific module in a pipeline-based Task-Oriented Dialog System.

## Domains:
{domain_str}

## Target Module:
{agent_type}({agent_role})

## Goal:
The strategies should aim to enhance the overall performance of the TODS by increasing task completion rate and reducing the average number of dialog turns (improving efficiency).

## Requirements for Strategies:
1. Each strategy must be a self-contained, actionable recommendation.
2. Describe the strategy concisely in 3 to 5 items.
3. Focus on specific techniques, architectural adjustments, or training methods relevant to the target module.
4. Explicitly address unique challenges or opportunities presented by the specified domain.
5. Implementation guidance must be clear enough for a developer to follow.
6. Optionally, include 0-3 few-shot examples if they perfectly illustrate the strategy's application

## Number of Strategies: {num}

## Output Format
Output MUST be a valid JSON array only, with no additional text, explanations, or markdown formatting.
[
  {
    "reason": "A clear, one sentence explanation of the performance bottleneck or optimization opportunity this strategy addresses for the specified module and domain.",
    "content": "The core strategy description and implementation steps (3-5 items). May include examples."
  },
  ...continue for all {num} strategies
]
\end{promptbox}

\subsubsection{Mutation Prompt}

\begin{promptbox}
## Task:
You are an expert in Task-Oriented Dialog Systems optimization. Based on the provided dialog data, analyze and optimize the strategy for the {agent_type} module.

## Context:
- Target Module: {agent_type}
- Agent Goal: {agent_goal}
- Domain: {domain_str}
- Dialog Result: {dialog_result}

## Input Data

### Dialog Goal
{goal}

### Dialog History
{formatted_history}

### Current Strategies
{strategies_by_type}

### Feedback Analysis
{evolve_data}

## TASK INSTRUCTIONS

### Step 1: Strategy Evaluation
Score the current strategy's effectiveness in this dialog:
- 1 (Helpful): Strategy contributed positively to dialog success or efficiency
- 0 (Neutral): Strategy had no clear positive or negative impact
- -1 (Harmful): Strategy directly contributed to dialog failure or inefficiency

Consider these factors when scoring:
1. Module specific performance in this dialog
2. Impact on overall task completion
3. Contribution to dialog efficiency (turn reduction)

### Step 2: Gap Analysis
Identify specific gaps between the current strategy and optimal performance by analyzing:
1. Dialog failures or inefficiencies in the history
2. Feedback insights and recommendations
3. Domain specific challenges that emerged

### Step 3: Strategy Optimization
Create an updated strategy that addresses the identified gaps while maintaining effective aspects of the current strategy. Ensure the updated strategy:
1. Addresses Specific Issues: Directly targets problems observed in the dialog
2. Provides Actionable Guidance: Clear, implementable recommendations
3. Leverages Domain Knowledge: Incorporates {domain_str} specific best practices
4. Balances Robustness and Efficiency: Maintains task completion while reducing unnecessary turns

### Step 4: Reasoning
Provide clear rationale explaining: What specific improvements the updated strategy makes

## Output Format
Output ONLY the JSON object. Do not include any additional text, explanations, or markdown formatting outside the JSON.
{
    "strategy": {
        "agent_type": "{agent_type}",
        "content": "Updated strategy description (1-3 few-shot expamles if needed)",
        "reason": "Clear rationale for the update one sentence explaining what issues were addressed)",
        "score": 1|0|-1
    }
}
\end{promptbox}

\subsubsection{Consolidation Prompt}
\begin{promptbox}
## Task:
Merge multiple semantically similar strategies into one comprehensive strategy for {agent_type} module for {domains_str} domain(s) in a Task-Oriented Dialog System.

## Merging Guidelines:
1. Analyze the provided strategies to identify: 1) Common themes and techniques, 2) Complementary ideas, 3) domain specific nuances.
2. Create a unified strategy that integrates the strongest elements from each original strategy, avoiding simple concatenation.
3. If strategies have conflicting advice, prioritize the approach that is most evidence based or best suited for the specified domain.
4. The merged strategy should be more generalizable than any single original strategy, while maintaining practical applicability.
5. Include 1-3 representative examples ONLY if they significantly enhance understanding of the merged approach. Adapt examples to better illustrate the integrated strategy.

{strategies_text}

# Output Format
Output ONLY a valid JSON object with exactly the structure below. Do not include any additional text, explanations, or markdown formatting.
{
    "content": "Merged strategy description here",
    "reason": "Summary of the merged strategy's purpose and value"
}
\end{promptbox}

\subsection{Manually Designed Strategies in Zero-shot Setup}
\label{app:manual_strategies}

\subsubsection{DST Agent Strategy}
\begin{promptbox}
1. Analyze User Utterance
- Extract slot value mentions from the user's current utterance
- Identify corrections, updates, or confirmations of existing values
- Analyze the previous User Output (user utterance). If there are any issues, provide critique. If the output is good, leave critique as empty string.

2. Update Belief State
- Only modify existing slots: DO NOT create new slots or domains
- Corrections: If user corrects a slot (e.g., "actually I want X"), replace the old value
- Updates: If user provides new information for a slot, update it
- Persistence: If slot not mentioned, keep its current value unchanged
- Handling uncertainty: If utterance is ambiguous, prefer keeping current value unless clear update

3. Quality Check
- Verify all slot values are consistent with the utterance
- Ensure domain constraints are respected
- Check for contradictions between slots
\end{promptbox}

\subsubsection{DP Agent Strategy}
\begin{promptbox}
1. First, analyze the belief state changes according to the lasted user utterance. If there are any issues, provide critique. If the output is good, leave critique as empty string.
2. Analyze the user's utterance and current belief state to determine the appropriate system action
3. CRITICAL: ALWAYS query the database for any information needed, NEVER use your own knowledge or common sense
4. Set "query_db" to true for ALL actions that require entity information, and specify the query parameters using the filled slots
5. Generate system action based ONLY on database query results, NEVER fabricate or assume any entity details, prices, addresses, or availability
6. Provide a reason for the DP output
\end{promptbox}

\subsubsection{NLG Agent Strategy}
\begin{promptbox}
1. First, analyze the previous Dialog Policy Module's output (system action). If there are any issues, provide critique. If the output is good, leave critique as empty string.
2. Understand the intent behind the system action. Use the provided strategies to handle specific response patterns.
3. Output your response in the specified JSON format with 'system_utterance' field and 'reason' field.
4. Ensure the response is natural, helpful, and appropriate for the dialog context. Keep the response concise but informative.
\end{promptbox}

\section{Case Study}
\label{app:case_study}

This section provides qualitative analyzes to empirically examine how DarwinTOD's evolutionary mechanisms operate in concrete dialog scenarios. Through detailed trajectory tracing and error case inspection, we aim to validate that the dual-loop architecture not only drives continuous strategy refinement but also effectively contains and recovers from potential cascading failures, key capabilities for lifelong learning in real-world deployments.

\subsection{Evolution Trajectory of a DP Agent Strategy}
\label{app:case_study_evolution}

This case study illustrates how an initial, generic strategy for the DP agent in the \textit{train} and \textit{restaurant} domains is progressively refined through DarwinTOD's evolutionary loop. The following snapshots show the strategy at four distinct generations.

\noindent \textbf{Initial Strategy (Generation 0):}
\begin{promptbox}
Augment the DP with a hierarchical policy structure. A top-level meta-policy selects a sub-policy (e.g., 'TicketBooking', 'MealSelection', 'SeatUpgrade') based on the user's intent and dialog stage. Each sub-policy is a specialized module for its sub-task. The state is shared hierarchically. For example, after booking a train, if the user asks 'Can I order a meal?', the meta-policy activates the 'MealSelection' sub-policy, which inherits the journey details (train number, date) without needing re-confirmation, enabling seamless multi-task dialogs.
\end{promptbox}

\noindent \textbf{10th-Generation Strategy:}
\begin{promptbox}
Implement a multi-stage, intent-aware proactive strategy with explicit state tracking. For train queries, separate information retrieval from booking: when user requests train details (departure time, travel time, train ID) without booking intent, provide all matching options immediately without requesting 'book people'. Only request booking-specific slots when user explicitly indicates booking intent (e.g., 'I want to book', 'reserve seats'). For restaurant searches, when user requests address/postcode before selection, automatically provide details for the first matching option with fallback option mention. Maintain a 'requested_slots' tracker to avoid redundant queries. 

Examples: 
1) User: 'train from A to B arriving by X' -> System queries DB, presents all matching trains with requested details, asks 'Would you like to book any of these?' 
2) User: 'address for restaurant' (before selection) -> System: 'The first option is X at address Y, postcode Z. Another option is...' 
3) When user provides all constraints for information request, skip booking slot requests until booking intent is confirmed.
\end{promptbox}

\noindent \textbf{20th-Generation Strategy:}
\begin{promptbox}
Implement a multi-domain dialog policy with intent-aware state tracking and parallel processing, enhanced with the following optimizations: 
1) **Intent-Aware Critical Slot Triggering with Explicit Intent Confirmation**: For restaurant domain, distinguish between search and booking intents. Only trigger booking-specific slot requests (book_people, book_time, book_day) after the user explicitly confirms booking intent by either selecting a specific restaurant or using booking-related language (e.g., 'book a table', 'reserve'). After providing restaurant recommendations, wait for user selection before requesting booking details. 
2) **Multi-Domain Parallel Query with Proactive Acknowledgment**: When users present requests for multiple domains simultaneously, immediately acknowledge both requests in the system response while processing queries in parallel. Use a unified response template: 'I'll help you with both [Domain1] and [Domain2]. For [Domain1], [specific action]. For [Domain2], [specific action].' This maintains user confidence in system understanding. 
3) **Domain-Specific Slot Requirement Optimization**: For train information queries (non-booking), require only {departure, destination, day, (leave_at OR arrive_by)}. Do not request book_people for information-only queries. For restaurant booking, require {name, book_time, book_day, book_people} only after explicit booking intent confirmation. 
4) **Proactive Constraint Verification**: Before executing database queries, verify that all mandatory constraints are logically consistent (e.g., book_time in appropriate format, day exists). If inconsistencies are detected, request clarification immediately rather than proceeding with query. 
5) **Response Efficiency Optimization**: When database queries return results for multiple domains, present information in a structured format: 'For your restaurant booking: [details]. For your train query: [details].' Include all user-requested attributes without requiring additional prompts. 

Examples: 
1) User: 'The Copper Kettle sounds good. Can you book a table for one person at 13:45 on Thursday? Also, I need a train from Cambridge to Kings Lynn on Friday, arriving by 18:15.' -> System: 'I'll help you with both requests. For the restaurant, I'll check availability at The Copper Kettle for 1 person at 13:45 on Thursday. For the train, I'll find options from Cambridge to Kings Lynn on Friday arriving by 18:15.' (Parallel queries triggered). 
2) User: 'I'm looking for British food, and I'd like it to be moderately priced.' -> System queries with {food: British, price_range: moderate, area: centre} and provides recommendations without requesting booking slots. 
3) User: 'What are the departure times and prices?' (for train) -> System queries with available constraints and provides information without requesting book_people.
\end{promptbox}

\noindent \textbf{Final Strategy (Generation 201):}
\begin{promptbox}
Implement a Unified, Proactive, and Result-Aware Multi-Domain Dialog Policy with Intent-Aware Phase Management, Parallel Query Execution, and Adaptive Constraint Handling. This integrated strategy for the DP module in train and restaurant domains is defined by below core principles: 
1) **Intent-Aware Phase Management & Critical Slot Triggering**: Classify user intent per domain (search/information-request vs. booking) using explicit cues and implicit context. Maintain explicit, phase-aware state machines per domain (e.g., QUERY, SELECT, BOOK). Define domain-specific mandatory slot sets that, when filled, immediately trigger a database query: restaurant search: {food, area, price_range}; train search: {departure, destination, day, (leave_at OR arrive_by)}; restaurant information request: {name}. Dynamic rule masks prevent premature requests for booking-phase slots (e.g., book_people, book_time, book_day) during query phases; booking slots are only requested after explicit booking intent confirmation (e.g., user selects a restaurant or uses booking-related language). 
2) **Parallel, Result-Aware Query Execution & Action Selection**: For multi-domain turns, execute database queries in parallel as soon as each domain's mandatory slots are filled. The system action (`system_action`) is determined **after** query execution based on the result set: if >1 matching entity -> `select` or `recommend`; if ==1 -> `inform` or `recommend`; if ==0 -> `nooffer`. This result-aware flow prevents pre-committing to actions like `request` for queries that cannot succeed. 
3) **Explicit Empty Result Handling & Proactive Constraint Relaxation**: Upon an empty query result (`nooffer`), immediately inform the user and proactively offer to relax constraints (e.g., 'I found no Chinese restaurants in the south. Would you like me to search in other areas or for other cuisines?'). This prevents repetitive slot requests for unsatisfiable queries. 
4) **Goal-Aware Proactive Information Delivery & Constraint Inference**: Leverage the full user goal context to pre-populate belief states for mandatory slots stated in the goal. Upon a successful query, autonomously provide all relevant informable slots for the first matching entity from the goal's 'request' list or user's explicit requests, requesting confirmation only if multiple viable options exist. Use derived information and intelligent defaults (e.g., for 'arrive_by' without 'leave_at', query with the earliest matching departure). 
5) **Adaptive Multi-Domain Handling with Context Carry-Over & Precise Clarification**: Process all active domains in parallel but independently. For sequential multi-domain requests, implement 'context carry-over': slots explicitly stated as shared (e.g., 'day', 'book_people') are inherited from the prior domain's confirmed belief state without re-requesting. When multiple slots need confirmation, generate precise multi-slot requests (e.g., `request(train.departure, train.travel_duration, restaurant.price_range, restaurant.area)`) instead of generic clarifications. For simultaneous multi-domain requests, immediately acknowledge both requests in the system response while processing queries in parallel, using a unified response template: 'I'll help you with both [Domain1] and [Domain2]. For [Domain1], [specific action]. For [Domain2], [specific action].' 

Examples: 
1) User: 'Find Italian food in centre, and a train to Cambridge Monday arriving by 18:00.'
 System classifies both as search intents, acknowledges both, executes parallel queries. If restaurant query returns 3 options, system uses `select()`; if train query returns 1, system uses `inform()`; if either returns 0, system uses `nooffer()` with relaxation offers. 
2) User: 'The Copper Kettle sounds good. Can you book a table for one person at 13:45 on Thursday? Also, I need a train from Cambridge to Kings Lynn on Friday, arriving by 18:15.'
 System: 'I'll help you with both requests. For the restaurant, I'll check availability at The Copper Kettle for 1 person at 13:45 on Thursday. For the train, I'll find options from Cambridge to Kings Lynn on Friday arriving by 18:15.'
 (Parallel queries triggered with explicit booking intent confirmation for restaurant). 
3) After a restaurant booking, user initiates a train request. System automatically inherits 'day' and 'book_people' from the restaurant context, querying the DB directly without re-requesting.
\end{promptbox}

\noindent \textbf{Analysis.} The evolutionary trajectory demonstrates a clear progression from a generic, hierarchical policy suggestion to a highly specialized, result-aware multi-domain strategy. Initially, the strategy proposes a modular but abstract architecture. Through interaction-driven feedback, it rapidly incorporates domain-specific distinctions (e.g., separating information queries from bookings) and introduces explicit tracking mechanisms. By the 20th generation, it evolves sophisticated features like parallel query acknowledgment, intent confirmation gates, and proactive constraint verification. The final strategy synthesizes these elements into a unified policy that dynamically couples query execution with result-aware action selection, implements intelligent constraint relaxation, and enables seamless cross-domain context transfer. This progression exemplifies how DarwinTOD's online execution and offline evolutionary dual-loop architecture, autonomously distills experiential feedback into increasingly precise, efficient, and robust dialog strategies, embodying the core principle of lifelong self-evolution.

\subsection{Strategy Evolution and Cascade Error Prevention}
\label{app:case_study_critique}

This section demonstrates how DarwinTOD's peer critique mechanism drives targeted strategy evolution and prevents cascading error propagation in the modular pipeline. We analyze a concrete dialog (ID: \texttt{multiwoz21-test-173}) spanning the \textit{taxi}, \textit{hotel}, and \textit{attraction} domains.

\noindent \textbf{Dialog Goal:}
\begin{promptbox}
You are planning your trip in Cambridge. You are looking for places to go in town. The attraction should be in the east. Make sure you get attraction type and postcode. You are also looking for a place to stay. The hotel should have a star of 4 and should include free wifi. The hotel should be in the same area as the attraction. Once you find the hotel you want to book it for 3 people and 4 nights starting from saturday. Make sure you get the reference number. You also want to book a taxi to commute between the two places. You want to leave the hotel by 09:45. Make sure you get contact number and car type.
\end{promptbox}

\noindent \textbf{User Initial Utterance:}
\begin{promptbox}
Are there any places in the east of town that are recommended to go to? Thanks.
\end{promptbox}

\noindent \textbf{DP Agent Strategy:}
\begin{promptbox}
Implement a 'Preference-Aware Proactive Selection' strategy with a fallback mechanism. 
1) **Preference Interpretation:** When a user expresses openness (e.g., 'open to any type'), treat this as an explicit instruction to proceed without further clarification. Do not request the slot. 
2) **Proactive Selection with Justification:** Immediately query the database with all other known constraints (e.g., area='centre'). From the results, select one entry (e.g., the first or most popular) and present it, explicitly stating the selection was made due to the user's openness (e.g., `inform('Since you are open to any type, I have selected the ADC Theatre.')`). Provide all requested slots (address, type, phone). 
3) **Multi-Goal Parallel Processing:** For multi-domain goals, process database queries in parallel where constraints are known. In this dialog, once 'hotel: area=centre' is confirmed at turn 2, immediately query for hotels matching 'area=centre, price range=cheap, stars=0, parking=yes' while simultaneously handling the attraction request. Present results as soon as available, interleaving domains to reduce sequential delays. 
4) **Constraint Validation Before Request:** Before requesting any slot (e.g., taxi departure), verify all prerequisite information is either already in the belief state or can be inferred (e.g., taxi departure can be inferred as the attraction's address once an attraction is selected). If not, prioritize completing the prerequisite domain first.
\end{promptbox}

Given the user query and the current belief state (attraction domain with \texttt{area='east'}), the DP agent genearates the following output:

\noindent \textbf{DP Agent Output:}
\begin{promptbox}
request(type)
\end{promptbox}

\noindent \textbf{DP Agent rationale:}
\begin{promptbox}
The user is asking for recommended places in the east of town, which corresponds to the attraction domain. The current belief state has 'area' filled as 'east' for attraction, but 'type' is still empty. To provide a recommendation, I need to know the type of attraction (e.g., museum, park, theatre) to query the database effectively. Therefore, I should request the 'type' slot from the user.
\end{promptbox}

The subsequent NLG agent, before generating the final response, critiqued this DP output:
\noindent \textbf{NLG Agent critique:}
\begin{promptbox}
The system action 'request(type)' is too vague and does not specify which domain's type is being requested, given the user's query involves attractions in the east of town. This could lead to confusion or an inefficient dialog turn.
\end{promptbox}

\noindent\textbf{Strategy Evolution:}
Based on this critique, the DP agent's strategy was evolved via the \textit{Mutation} operator. The updated strategy incorporates explicit domain specification to eliminate ambiguity in multi-domain contexts:

\begin{promptbox}
Implement an 'Enhanced Domain-Aware Proactive Selection' strategy with the following updates:
1. Domain-Specific Slot Requests: Always specify the domain when requesting slots in multi-domain dialogs. For example, instead of request(type), use request(attraction.type) or natural language like 'What type of attraction are you interested in?'. This applies to all requests (e.g., for hotel: 'request(hotel.stars)', for taxi: 'request(taxi.departure_time)').
2. Preference Interpretation with Domain Context: When a user expresses openness (e.g., 'open to any type'), interpret it within the current domain focus. If the openness is ambiguous, confirm the domain first (e.g., 'Are you open to any attraction type?') before proceeding.
3. Proactive Selection with Justification and Domain Tagging: When selecting an entry due to user openness, present it with explicit domain tagging (e.g., inform(attraction='ADC Theatre', type='theatre', domain='attraction') and state the justification. Provide all requested slots with domain context.
4. Multi-Goal Parallel Processing with Domain Clarity: Process queries in parallel, but when presenting results, interleave domains with clear labels (e.g., 'For attractions: ... For hotels: ...'). Ensure that dependency constraints (e.g., hotel area matching attraction area) are validated before presenting options.
5. Constraint Validation Before Request with Prerequisite Checks: Before requesting any slot, verify prerequisites are met and infer where possible (e.g., taxi departure can be inferred from hotel address). If prerequisites involve another domain, prioritize completing that domain first, and use domain-specific requests to avoid confusion.

Few-shot examples:
- Example 1: In a multi-domain context, if user asks for 'places in the east', system should request as 'request(attraction.type)' or say 'What type of attraction, like a museum or park, are you looking for?'.
- Example 2: When user is open to any hotel type, system queries hotels with known constraints, selects one, and informs: 'Since you are open to any hotel type, I have selected the Hilton Hotel with 4 stars and free wifi. The address is...'.
- Example 3: For taxi booking, after hotel is booked, system infers departure as hotel address and requests 'taxi.contact_number' and 'taxi.car_type' specifically.
\end{promptbox}

\noindent\textbf{Evolving Rationale:}
\begin{promptbox}
This update directly addresses the vagueness in slot requests by enforcing domain specification, reducing confusion and inefficiency in multi-domain dialogs, while maintaining proactive selection and parallel processing for robustness.
\end{promptbox}

\noindent\textbf{Preventing Cascading Error Propagation:}
Despite the ambiguous DP action (\texttt{request(type)}), the NLG agent's critique and corrective response prevented a suboptimal user experience. The NLG agent generated the following natural language response and rationale:
\begin{promptbox}
Sure, I can help you find recommended places in the east of town. To narrow it down, what type of attraction are you interested in, such as a museum, park, or something else?
\end{promptbox}
\begin{promptbox}
The user asked for recommended places in the east of town, which aligns with the attraction domain. The system action should have been more specific, like 'request(attraction_type)', but I interpreted it as requesting the type of attraction to provide relevant recommendations. I kept the response natural and helpful by acknowledging the request and asking for clarification on the attraction type to assist effectively.
\end{promptbox}

\noindent\textbf{Analysis.} This case illustrates the synergistic operation of DarwinTOD's dual-loop architecture. The online peer critique mechanism immediately identified a vague, domain ambiguous action that could have led to user confusion or extended clarification loops. By generating a corrective, domain specific response, the NLG agent contained the potential error within the current turn, preventing its propagation through subsequent dialog states. Offline, this critique triggered a targeted mutation of the DP strategy, which now explicitly mandates domain qualified slot requests in multi-domain contexts. This evolution not only fixes the specific flaw but also generalizes the improvement to future dialogs. The episode demonstrates how the pipeline's inherent modularity, combined with inter-agent critique, transforms a potential cascading errors into a strength: each module acts as a semantic firewall, while the evolutionary loop accumulates these local corrections into globally robust strategies.

\end{document}